\documentclass[10pt,twocolumn]{IEEEtran}

\usepackage{amsmath,amsfonts,graphics,amssymb,epsfig,subfigure,color,cite}
\usepackage{amsthm,textcomp}
\theoremstyle{plain}
\newtheoremstyle{mystyle}
  {0mm}
  {0mm}
  {}
  {4mm}
  {\bfseries}
  {:}
  { }
  {\thmname{#1}\thmnumber{ #2}\thmnote{ (#3)}}
\theoremstyle{mystyle}

\usepackage{stackengine}
\newcommand\xrowht[2][0]{\addstackgap[.5\dimexpr#2\relax]{\vphantom{#1}}}
\usepackage{array}
\usepackage{multirow}
\usepackage{enumerate}
\usepackage{algorithm}
\usepackage{algpseudocode}
\usepackage{algcompatible}
\algblockdefx{FORP}{ENDFORP}[1]%
  {\textbf{for }#1 \textbf{do in parallel}}%
  {\textbf{end}}
\algblockdefx{FOR}{ENDFOR}[1]%
  {\textbf{for }#1 \textbf{do}}%
  {\textbf{end}}
\algblockdefx{IF}{ENDIF}[1]%
  {\textbf{if }#1 \textbf{then}}%
  {\textbf{end}}
\algnewcommand\algorithmicprocedure{\textbf{function}}
\algnewcommand\FUNC{\item[\algorithmicprocedure]}%
\algnewcommand\algorithmicendprocedure{\textbf{end function}}
\algnewcommand\ENDFUNC{\item[\algorithmicendprocedure]}%
 \usepackage{setspace}
\let\Algorithm\algorithm
\renewcommand\algorithm[1][]{\Algorithm[#1]\setstretch{1.4}}
\usepackage{float}
\usepackage{makecell}
\usepackage{graphicx}
\usepackage{lipsum}

\usepackage[caption=false,font=normalsize,labelfont=sf,textfont=sf]{subfig}

\newtheorem{thm}{Theorem}
\newtheorem{lem}{Lemma}

\floatname{algorithm}{Procedure}

\newcommand{\argmin}{\operatornamewithlimits{argmin}}
\makeatletter
\newcommand{\vast}{\bBigg@{4.5}}
\newcommand{\Vast}{\bBigg@{7.5}}
\makeatother

\allowdisplaybreaks

\begin{document}
    \title{FedVQCS: Federated Learning via \\ Vector Quantized Compressed Sensing}
	\author{Yongjeong Oh, \IEEEmembership{Student Member,~IEEE}, Yo-Seb Jeon, \IEEEmembership{Member,~IEEE}, \\Mingzhe Chen, \IEEEmembership{Member,~IEEE}, and Walid Saad, \IEEEmembership{Fellow,~IEEE}
	    \thanks{Yongjeong Oh is with the Department of Electrical Engineering, POSTECH, Pohang, Gyeongbuk 37673, Republic of Korea (e-mail: yongjeongoh@postech.ac.kr).}
        \thanks{Yo-Seb Jeon is with the Department of Electrical Engineering, POSTECH, Pohang, Gyeongbuk 37673, Republic of Korea. He is also with the Institute for Convergence Research and Education in Advanced Technology, Yonsei University, Incheon 21983, Republic of Korea (e-mail: yoseb.jeon@postech.ac.kr).}
		\thanks{Mingzhe Chen is with the Department of Electrical and Computer Engineering and Institute for Data Science and Computing, University of Miami, Coral Gables, FL 33146, USA (e-mail: mingzhe.chen@miami.edu).}
		\thanks{Walid Saad is with the Wireless@VT, Bradley Department of Electrical and Computer Engineering, Virginia Tech, Arlington, VA 22203, USA (e-mail: walids@vt.edu).}
	}
	\vspace{-2mm}	
	
	\maketitle
	\vspace{-12mm}

	\begin{abstract} 
	    In this paper, a new communication-efficient federated learning (FL) framework is proposed, inspired by vector quantized compressed sensing. The basic strategy of the proposed framework is to compress the local model update at each device by applying dimensionality reduction followed by vector quantization. Subsequently, the global model update is reconstructed at a parameter server by applying a sparse signal recovery algorithm to the aggregation of the compressed local model updates. By harnessing the benefits of both dimensionality reduction and vector quantization, the proposed framework effectively reduces the communication overhead of local update transmissions. Both the design of the vector quantizer and the key parameters for the compression are optimized so as to minimize the reconstruction error of the global model update under the constraint of wireless link capacity. By considering the reconstruction error, the convergence rate of the proposed framework is also analyzed for a non-convex loss function. Simulation results on the MNIST and FEMNIST datasets demonstrate that the proposed framework provides more than a 2.4$\%$ increase in classification accuracy compared to state-of-the-art FL frameworks when the communication overhead of the local model update transmission is 0.1 bit per local model entry.

	\end{abstract}


	\section{Introduction}\label{Sec:Intro}
	Federated learning (FL) is a distributed machine learning technique which aims at training a {\em global} model on a parameter server (PS) by collaborating with distributed wireless devices, each with its own local training dataset \cite{Konecny:15,McMahan:17}.
    In a typical FL framework, each device updates its {\em local} model based on the local training dataset and then sends a local model update to the PS \cite{McMahan:17}.
    After the transmission from the devices, the PS updates its global model by aggregating the local model updates sent by the devices and then distributes the updated global model to the devices. 
    This two-step training process continues until the global model at the PS converges. 
    The above FL framework can help preserve the privacy of the data generated by the devices.
    Hence, it has drawn great attention as a privacy-preserving alternative to centralized machine learning \cite{Niknam:19,Zhu:20,Gunduz:20,Samarakoon:19,Chen:21}.
	While FL offers an important benefit in terms of data privacy, bringing this technique into real-world applications also faces several challenges.
	One major such challenge is the significant communication overhead required for transmitting the local model updates from the wireless devices to the PS because the wireless links connecting them may have limited capacity in practical communication systems.
    This challenge becomes critical when the dimensionality of the local model updates is much larger than the capacity of the wireless links.  
    For example, when the wireless link capacity is $20$ Mbps and the global model is ResNet-$18$ in \cite{ResNet} with $1.1\times 10^7$ trainable parameters, transmitting the local model update for $100$ times with $10$ devices takes about $1.8\times 10^4$ seconds.
    Such long communication times may not be acceptable for low-power devices and practical FL applications that often must operate at much lower latency.


    \subsection{Prior Works}
	To address the above challenge, communication-efficient FL via local model update compression was extensively studied. 
	The basic idea in these prior works is to apply {\em lossy} compression to the local model updates, in order to reduce the communication overhead required for transmitting these updates. 
	A representative approach in this direction is the scalar quantization approach studied in \cite{Konecny:16,Alistarh:17,SignSGD,Bennis:21}, in which the entries of the local model update are independently quantized by a scalar quantizer. 
	By extending this approach, a vector quantization approach for communication-efficient FL  was also studied in \cite{UVeqFed,UVeqFed2,Du:20}. 
	In this approach, a partition of each local model update is quantized by a vector quantizer.
    The common limitation of the {\em quantization-only} approach, however, is that significant quantization error is inevitable when the communication overhead of the local model update transmission must be less than one bit per local model entry.

    Recently, to overcome the limitation of the quantization-only approach, communication-efficient FL via quantized compressed sensing (QCS) has been explored in \cite{QCS1,QCS2,FedQCS}. 
    A key motivation in this approach is the sparsity of the local model update, obtained either naturally or by applying sparsification  \cite{Aji:17,Wangni:18,Lin:18,Amiri:TSP,Amiri:TWC,Jeon:21}.
	To exploit this sparsity, the basic premise of  the QCS approach is to project the local model updates onto a lower dimensional space as in compressed sensing (CS) before they are quantized. 
    Because of the ensuing dimensionality reduction, this approach can further reduce the communication overhead of the local update transmissions when compared with the quantization-only approach.
	A QCS-based compression method with one-bit quantization was studied in \cite{QCS1} and \cite{QCS2}, in which the entries of the local model update after dimensionality reduction were independently quantized by a one-bit scalar quantizer.  
	This method was extended to the compression with multi-bit quantization in \cite{FedQCS} which allows greater flexibility regarding the number of scalar quantization bits.
	The work in \cite{FedQCS} demonstrated that the use of multi-bit quantization provides a better tradeoff between communication efficiency and FL performance when compared with the one-bit quantization \cite{FedQCS}. 
    The existing QCS-based compression methods, however, only consider scalar quantization which is generally inferior to vector quantization in terms of quantization error \cite{GainShape}; thereby, the compression error of these methods becomes problematic as the level of the compression increases.
    To our best knowledge, local model update compression that harnesses the benefits of both vector quantization and dimensionality reduction has never been studied before, despite its potential for improving the communication efficiency of FL and mitigating the compression error of the local model updates.


	\subsection{Contributions}
	The main contribution of this paper is a novel communication-efficient FL framework that leverages, for the first time, both vector quantization and CS-based dimensionality reduction to compress local model updates at the wireless devices.
    We also propose a promising strategy for reconstructing the global model update at the PS from the compressed local model updates.
    More importantly, we minimize the reconstruction error at the PS by optimizing both the design of the vector quantizer and the key parameters for the compression and, then, we analyze the convergence rate of our framework with consideration of the reconstruction error. 
    In summary, our major contributions include the following:
	\begin{itemize}
		\item We present a communication-efficient FL framework via vector quantized compressed sensing (VQCS), dubbed {\em FedVQCS}.
		In this framework, we compress each local model update by reducing its dimensionality  based on CS and, then, quantizing a low-dimensional local model update by using a vector quantizer. 
		For accurate but efficient reconstruction of the global model update at the PS, we aggregate a group of the compressed local model updates and, then, estimate the aggregated model update by applying a sparse signal recovery algorithm. 
		By harnessing the benefits of both dimensionality reduction and vector quantization, FedVQCS significantly reduces the communication overhead required for transmitting the local model updates at the wireless devices.

		\item We optimize the design of the vector quantizer used for the local update compression strategy of FedVQCS.
	    A key feature of our compression strategy is that the local model update after the CS-based dimensionality reduction can be modeled as an independent and identically distributed (IID) Gaussian random vector by the central limit theorem. 
		By utilizing this feature, we derive the optimal shape and gain quantizers for an IID Gaussian random vector based on the Lloyd--Max algorithm \cite{LloydMax} and  Grassmanian codebook \cite{Grassline}, respectively.
		We also derive the optimal bit allocation between the shape and gain quantizers by characterizing the mean-squared-error (MSE) performance of these quantizers.

		\item We optimize the key design parameters of the local update compression strategy of FedVQCS, that include: (i) the number of quantization bits, (ii) sparsity level, and (iii)  dimensionality reduction ratio. 
		Unlike previous works where these parameters were heuristically determined as hyper-parameters, we analytically determine the best set of parameters for minimizing the reconstruction error of the global model update at the PS under a capacity constraint.
        A prominent feature of our parameter optimization is that it can be performed locally at each device without information exchange among the devices.

		\item {We analyze the convergence rate of FedVQCS, taking into account all errors that should be considered in the QCS-based FL framework, including sparsification errors of the local model update at the devices and reconstruction errors of the global model update at the PS.}
		Our analysis demonstrates that FedVQCS is guaranteed to converge to a stationary point of a smooth loss function at the rate of $\mathcal{O}\big(\frac{1}{\sqrt{T}}\big)$, where $T$ is the number of total iterations of the stochastic gradient descent (SGD) algorithm.   
		Our analysis also shows that the upper bound of the convergence rate decreases as the MSE of the global model update reconstructed at the PS reduces, implying that our parameter optimization contributes to improving the convergence rate of FL.

		\item Using simulations, we demonstrate the superiority of FedVQCS over state-of-the-art FL frameworks for an image classification task using the MNIST dataset \cite{MNIST} and the FEMNIST dataset \cite{FEMNIST}.
		Our simulation results demonstrate that FedVQCS provides more than a 2.4$\%$ increase in classification accuracy compared to the state-of-the-art FL frameworks when the communication overhead of the local model update transmission is 0.1 bit per local model entry.
		Meanwhile, the results on the MNIST dataset demonstrate that FedVQCS with a 0.1-bit overhead per local model entry yields only a 2$\%$ decrease in classification accuracy when compared with the case that performs no local update compression.
		The effectiveness of the parameter optimization of FedVQCS is also verified in terms of classification accuracy.

	\end{itemize}

	In our preliminary work \cite{GC paper}, we introduced a communication-efficient FL framework based on VQCS including the optimal design of the vector quantizer. We extend this preliminary work by making important progress on improving and analyzing the VQCS-based FL framework. First of all, we newly provide the parameter optimization for the local model update compression, which is critical for minimizing the reconstruction error of the global model update at the PS, as will be demonstrated in Sec. V.
	Furthermore, we present the convergence analysis of the VQCS-based FL framework to theoretically justify the effectiveness of this framework, which was not explored in \cite{GC paper}.

    The remainder of the paper is organized as follows. 
    In Sec.~II, we first introduce a wireless FL system considered in the paper. We then present a typical FL framework and its challenge. 
    In Sec.~III, we present the proposed FedVQCS framework, which addresses the above challenge by leveraging the idea of VQCS. In Sec.~IV, we optimize a quantizer design and key parameters of FedVQCS and analyze its convergence rate. In Sec. V, we provide simulation results to verify the superiority of the proposed framework. Finally, in Sec. VI, we present our conclusions and future research directions.

	\section{System Model}\label{Sec:System}

    Consider a wireless FL system in which a PS trains a global model by collaborating with a set $\mathcal{K}$ of $K$ wireless devices over wireless links with limited capacity.
    The data samples for training the global model are assumed to be distributed over the wireless devices only, while the PS does not have direct access to them.
    Let $\mathcal{D}_k$ be the {\em local training dataset} available at device $k\in \mathcal{K}$.
    The global model at the PS is assumed to be represented by a parameter vector $\boldsymbol{w}\in \mathbb{R}^{\bar{N}}$, where $\bar{N}$ is the number of global model parameters.  
    The goal of FL is to find the best parameter vector that minimizes the {\em global} loss function, defined as 
	\begin{align}\label{eq:global_loss0}
	    F(\boldsymbol{w}) = \frac{1}{|\mathcal{D}|} 
	    \sum_{\boldsymbol{u}\in \mathcal{D}} f(\boldsymbol{w};\boldsymbol{u}),
	\end{align}
	where $f(\boldsymbol{w};\boldsymbol{u})$ is a loss function that measures how well the global model with the parameter vector $\boldsymbol{w}$ fits one particular data sample $\boldsymbol{u}\in\mathcal{D}_k$, and $\mathcal{D} = \cup_k \mathcal{D}_k$.
    The global loss function in \eqref{eq:global_loss0} can be rewritten as 
	\begin{align}\label{eq:global_loss}
	    F(\boldsymbol{w}) = \frac{1}{\sum_{j=1}^K|\mathcal{D}_j|} \sum_{k=1}^K  |\mathcal{D}_k|  F_k(\boldsymbol{w}),
	\end{align}
	where $F_k(\boldsymbol{w})=\frac{1}{|\mathcal{D}_k|} \sum_{\boldsymbol{u}\in\mathcal{D}_k} f(\boldsymbol{w};\boldsymbol{u})$ is a {\em local} loss function at device $k$.
    Key symbols and their definitions considered in the paper are summarized in Table~\ref{table:param}.

    \begin{table}[t]\vspace{-1mm}
        \renewcommand{\arraystretch}{1.2}
        \centering
        \small
        \caption{Key symbols and their definitions} \vspace{-1mm} 
        \label{table:param}
        \begin{tabular}{|c| c|}
            \hline
            \bfseries Symbol & \bfseries  Definition\\
            \hline \hline
            $K$ & The number of devices \\ \hline 
            $G$ & The number of groups \\ \hline
            $B$ & The number of blocks \\ \hline 
            $N$ & Dimension of local block update \\ \hline
            $R_k$  & Dimensionality reduction ratio \\ \hline
            $Q_k$  & The number of quantization bits  \\ \hline
            $S_k$ & Sparsity level \\ \hline
            $M_k$ & Dimension of a compressed local block update 
            \\ \hline
            $\boldsymbol{g}^{(t)}_k$ & Local model update \\ \hline
            ${\boldsymbol g}_k^{(t,b)}$ & Local block update \\ \hline
            $\tilde{\boldsymbol g}_{k}^{(t,b)}$ & Local block update after the sparsification 
            \\ \hline
            ${\boldsymbol \Delta}_k^{(t,b)}$ & Residual vector \\ \hline
            $\boldsymbol{g}_{\mathcal{K}}^{(t)}$ & Global model update \\ \hline
            $\boldsymbol{g}_{\mathcal{K}}^{(t,b)}$ & Global block update \\ \hline
            ${\boldsymbol A}_{R_k}$ & Projection matrix \\ \hline
            $\alpha_k^{(t,b)}$ & Scaling factor \\ \hline
            ${\boldsymbol x}_{k}^{(t,b)}$ & Compressed local block update \\ \hline
            $\hat{\boldsymbol x}_{k}^{(t,b)}$ & Quantized local block update \\ \hline
            ${\boldsymbol v}_{k,p}^{(t,b)}$ & Sub-vector of a compressed local block update
            \\ \hline\xrowht{16pt}
            $\hat{\boldsymbol v}_{k,p}^{(t,b)}$ & 
            \makecell{Quantized sub-vector of \\  a compressed local block update}
            \\ \hline
            $\mathcal{K}_g$ & Index set of the devices in the $g$-th group 
            \\ \hline\xrowht{16pt}
            ${\boldsymbol y}_{\mathcal{K}_g}^{(t,b)}$ & \makecell{Aggregation of the quantized \\ local block updates for the group $g$} 
            \\ \hline\xrowht{16pt}
            $\tilde{\boldsymbol g}_{\mathcal{K}_g}^{(t,b)}$ & \makecell{Aggregation of the local  \\block updates for the $g$-th group} \\ \hline
            $\hat{\boldsymbol g}_{\mathcal{K}_g}^{(t,b)}$ & Estimate of $\tilde{\boldsymbol g}_{\mathcal{K}_g}^{(t,b)}$ \\ \hline
            $\hat{\boldsymbol g}_{\mathcal{K}}^{(t,b)}$ & Estimate of ${\boldsymbol g}_{\mathcal{K}}^{(t,b)}$ \\ \hline
            $\hat{\boldsymbol g}_{\mathcal{K}}^{(t)}$ & Estimate of ${\boldsymbol g}_{\mathcal{K}}^{(t)}$ \\ \hline
        \end{tabular}\vspace{-3mm}
    \end{table}

    \subsection{A Typical FL Framework}\label{Sec:FedAvg}
    A typical FL approach for minimizing the global loss function in \eqref{eq:global_loss} involves alternating between {\em local} model update at wireless devices and {\em global} model update at the PS in each communication round, as explained in \cite{McMahan:17}. 

    \subsubsection{Local model update at wireless devices} 
    In the local model update process, each wireless device updates a local parameter vector based on its own local training dataset. Then each device sends a local model update (i.e., the difference between the parameter vectors before and after the local update) to the PS.
	Let $\boldsymbol{w}^{(t)}\in{\mathbb R}^{\bar{N}}$ be the parameter vector shared by the devices at communication round $t \in \{1,\ldots, T\}$, where $T$ is the total number of communication rounds.
	Assume that every device employs a mini-batch SGD algorithm with $E\geq 1$ local iterations for updating the parameter vector $\boldsymbol{w}^{(t)}$.
	Then the updated parameter vector at device $k$ after $E\geq 1$ local iterations is given by
	\begin{align}
        \boldsymbol{w}^{(t,e+1)}_k = \boldsymbol{w}^{(t,e)}_k - \eta^{(t)}\nabla F_{k}^{(t,e)}\big(\boldsymbol{w}^{(t,e)}_k \big),
	\end{align}
	for all $e \in \{1,\ldots,E\}$, where $\boldsymbol{w}^{(t,1)}_k = \boldsymbol{w}^{(t)}$, $\eta^{(t)}$ is a local learning rate, 
    \begin{align}\label{eq:local_grad0}
	    \nabla F_{k}^{(t,e)}\big(\boldsymbol{w}^{(t,e)}_k\big) = \frac{1}{|\mathcal{D}_k^{(t,e)}|} \sum_{\boldsymbol{u}\in\mathcal{D}_k^{(t,e)}} \nabla f(\boldsymbol{w}^{(t,e)}_k;\boldsymbol{u}),
	\end{align}
	and $\mathcal{D}_k^{(t,e)}$ is a mini-batch randomly drawn from $\mathcal{D}_k$ at the $e$-th local iteration of round $t$.
	As a result, the local model update sent by device $k$ in round $t$ is determined as 
	\begin{align}
        \boldsymbol{g}^{(t)}_k = \frac{1}{\eta^{(t)}E}(\boldsymbol{w}^{(t)}-\boldsymbol{w}^{(t,E+1)}_k) \in \mathbb{R}^{\bar{N}}.
	\end{align}

    \begin{figure*}[t!]
        \centering
        {\epsfig{file=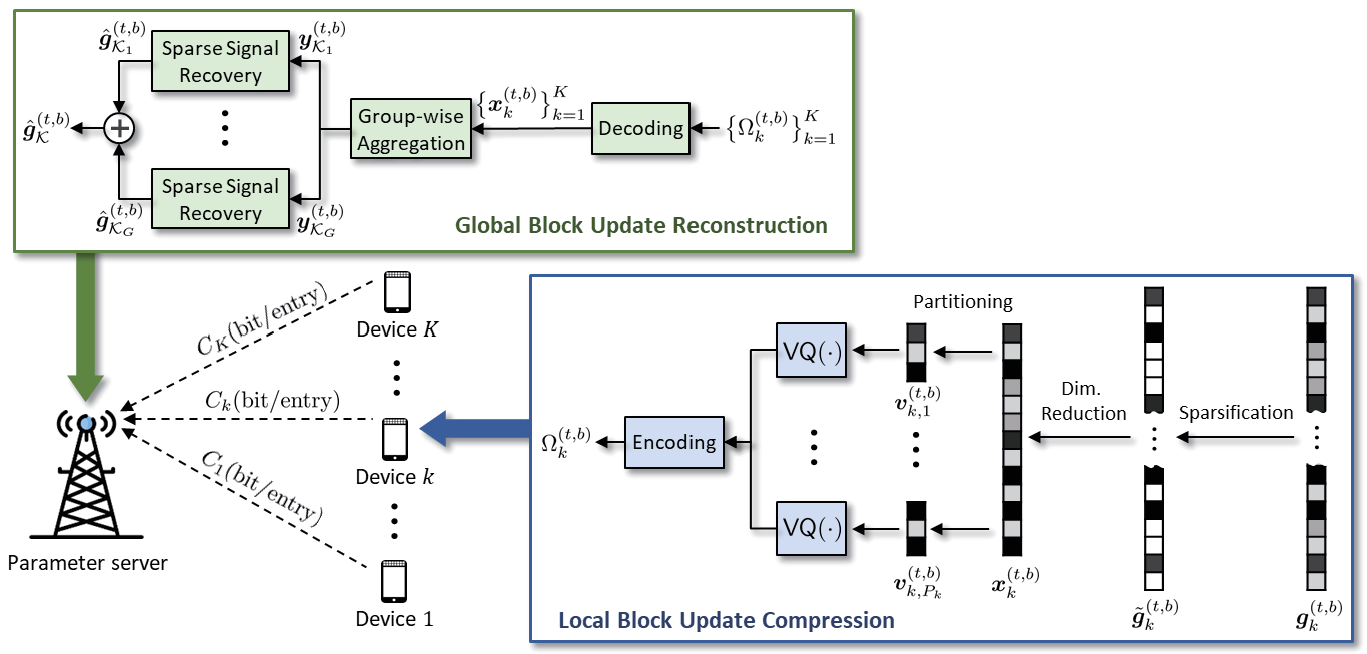, width=15cm}}
        \caption{An illustration of a single communication round $t$ within the proposed FL framework when deployed over a wireless network.}  
        \label{fig:System}
    \end{figure*}
        
    \subsubsection{Global model update at the PS} 
    In the global model update process, the PS updates a global parameter vector by aggregating the local model updates sent by $K$ devices. 
    Then the PS broadcasts the updated parameter vector to these devices.
    Under the assumption of perfect reception of $K$ local model updates, the PS can reconstruct the {\em global} model update, defined as
    \begin{align}\label{eq:global_model_update_true}
	    \boldsymbol{g}_{\mathcal{K}}^{(t)} = \sum_{k=1}^K \rho_k\boldsymbol{g}_{k}^{(t)},
	\end{align}
	where $\rho_k \triangleq \frac{\sum_{e=1}^{E}|\mathcal{D}_k^{(t,e)}|}{\sum_{j=1}^K \sum_{e=1}^{E}|\mathcal{D}_j^{(t,e)}|}$ is invariant in each communication round. 
	If the PS employs the SGD algorithm for updating the parameter vector $\boldsymbol{w}^{(t)}$, then the updated parameter vector in round $t$ will be given by 
	\begin{align}\label{eq:global_update_true}
	    \boldsymbol{w}^{(t+1)} = \boldsymbol{w}^{(t)} - \gamma^{(t)}\boldsymbol{g}_{\mathcal{K}}^{(t)},
	\end{align}
	where $\gamma^{(t)}$ is a global learning rate.
	Then the PS broadcasts $\boldsymbol{w}^{(t+1)}$ to the wireless devices, which triggers the start of the local model update process in round $t+1$.

	\subsection{Key Challenge in Wireless FL}
	A major challenge in realizing the FL framework in Sec.~\ref{Sec:FedAvg} is the significant communication overhead required for transmitting the local model updates from the wireless devices to the PS because the wireless links connecting them may have limited capacity in practical communication systems.
	This challenge can be mitigated by applying lossy compression to the local model updates, but such a compression approach leads to an inevitable error in the global model update reconstructed at the PS.
	Moreover, in general, the higher the level of compression at the wireless devices, the larger the reconstruction error at the PS. 
	Therefore, it is essential to develop a rigorous compression strategy for local model updates that provides a considerable reduction in the communication overhead while also minimizing the reconstruction error at the PS.

	\section{Proposed FedVQCS framework}\label{Sec:FedVQCS}
	In this section, we present a communication-efficient FL framework, called FedVQCS, that can reduce the communication overhead for transmitting local model updates at the wireless devices while also enabling an accurate reconstruction of the global model update at the PS.

    \subsection{Overview of FedVQCS}
    The basic strategy of FedVQCS is to compress the local model updates at the wireless devices by leveraging both vector quantization and CS-based dimensionality reduction while reconstructing the global model update at the PS via sparse signal recovery.
    The high-level procedure of FedVQCS is illustrated in Fig.~\ref{fig:System} and described below.

    \subsubsection{Compression of the local model updates} 
    In each communication round, each device divides its local model update into $B$\footnote{The number of blocks $B$ is pre-determined at the PS before the beginning of the training process according to the computational power of the PS.} blocks, each of which has a length of $N = \frac{\bar{N}}{B}$\footnote{In practice, this process can be realized by randomly shuffling local model update entries and then dividing blocks in order. The information about random shuffle can be expressed as a random seed number that can be shared between the PS and devices before the beginning of the training process.}. 
    Let $\boldsymbol{g}_{k}^{(t,b)} \in \mathbb{R}^N$ be the $b$-th block of $\boldsymbol{g}_{k}^{(t)}$, and we refer to this vector as the $b$-th local {\em block} update of device $k$ in round $t$. 
    Each device compresses these $B$ local block updates in parallel by applying a local block compression process denoted by a function ${\sf BlkComp}(\cdot)$.
    The motivation behind this {\em block-wise} compression is to reduce the computational complexity for reconstructing the global model update at the PS, as considered in \cite{FedQCS}.
    The compressed local block updates are encoded into digital bits and then transmitted to the PS through wireless links.  

    \subsubsection{Reconstruction of the global model update} 
    In each communication round, the PS reconstructs a global model update from the compressed local model updates sent by the wireless devices.
    Let $\boldsymbol{g}_{\mathcal{K}}^{(t,b)} = \sum_{k=1}^K \rho_k \boldsymbol{g}_{k}^{(t,b)}$ be the $b$-th global block update in round $t$ which is the aggregation of the local block updates sent by $K$ devices. 
    The PS first reconstructs $\boldsymbol{g}_{\mathcal{K}}^{(t,b)}$ for every block $b \in \{1,\ldots,B\}$, by applying a global block reconstruction process denoted by a function ${\sf BlkReconst}(\cdot)$. 
    The PS then concatenates the reconstructed block updates to construct the \textit{global} model update.
    This process is expressed as $\hat{\boldsymbol{g}}_{\mathcal{K}}^{(t)} = {\sf Concatenate}(\{\hat{\boldsymbol{g}}_{\mathcal{K}}^{(t,b)}\}_{b=1}^B)$, where $\hat{\boldsymbol{g}}_{\mathcal{K}}^{(t,b)}$ is the reconstruction of ${\boldsymbol g}_{\mathcal{K}}^{(t,b)}$ determined by the global block reconstruction process. 

    \setlength{\textfloatsep}{15pt}
    \begin{algorithm}[t]
        \caption{Federated Learning via Vector Quantized Compressed Sensing (FedVQCS)}\label{alg:FedVQCS}
        {\small
        {\begin{algorithmic}[1]
            \REQUIRE ${\boldsymbol w}^{(1)}$,  $\{\mathcal{K}_g\}_{g=1}^G$, $\{C_k\}_{k=1}^K$
            \ENSURE ${\boldsymbol w}^{(T)}$
            \FOR {$t=1$ to $T$}
                \STATE \!\!\!{\em At the wireless devices:}
                \FORP {Each device $k\in\mathcal{K}$}
                    \STATE ${\boldsymbol g}_k^{(t)} = {\sf LocalUpdate}({\boldsymbol w}^{(t)}, \mathcal{D}_k)$
                    \FORP {Each block $b\in\mathcal{B}$}
                         \STATE \!\!\!\!$\big\{\Omega_{k}^{(t,b)}\!,{\boldsymbol \Delta}_{k}^{(t,b)}\big\} = {\sf BlkComp}\big({\boldsymbol g}_k^{(t,b)}\!,{\boldsymbol \Delta}_{k}^{(t-1,b)}\big)$
                    \ENDFORP
                    \STATE Transmit $\{\Omega_k^{(t,b)}\}_{b=1}^{B}$ to the PS.
                \ENDFORP
                \STATE \!\!\!{\em At the parameter server:}
                \FORP {Each block $b\in\mathcal{B}$}
                    \STATE $\hat{\boldsymbol g}_{\mathcal{K}}^{(t,b)} = {\sf BlkReconst}\big(\{\Omega_{k}^{(t,b)}\}_{k=1}^K\big)$
                \ENDFORP
                \STATE $\hat{\boldsymbol g}_{\mathcal{K}}^{(t)} = {\sf Concatenate}\big(\{\hat{\boldsymbol g}_{\mathcal{K}}^{(t,b)}\}_{b=1}^B\big)$
                \STATE ${\boldsymbol w}^{(t+1)} = {\boldsymbol w}^{(t)} - \gamma^{(t)} \hat{\boldsymbol g}_{\mathcal{K}}^{(t)}$
                \STATE Broadcast ${\boldsymbol w}^{(t+1)}$ to the wireless devices.
            \ENDFOR
        \end{algorithmic}}}
    \end{algorithm}
    \begin{figure}[t]
    	\centering
    	{\epsfig{file=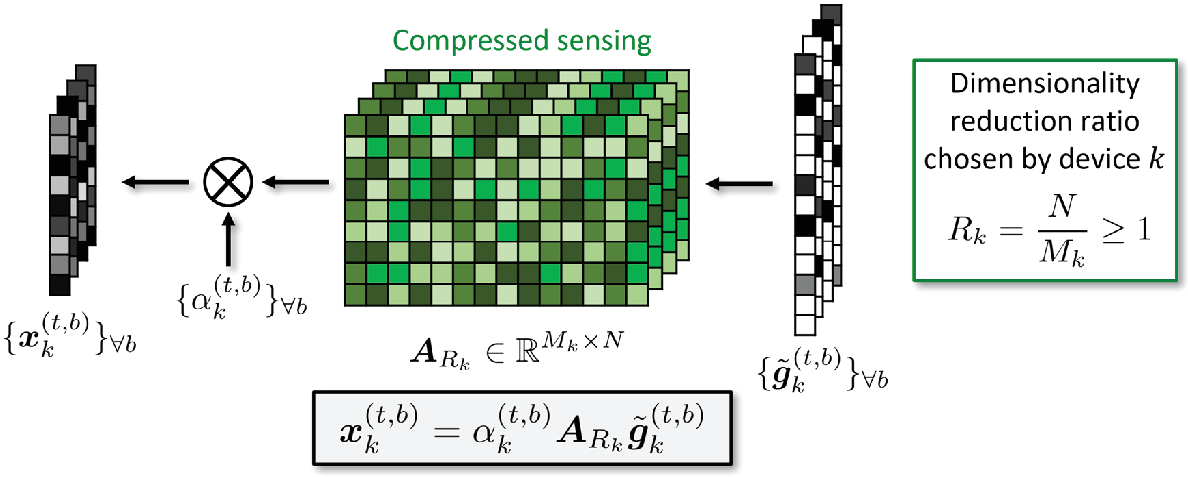, width=8.8cm}} 
    	\caption{Dimensionality reduction step of the proposed FedVQCS framework.}  
    	\label{fig:DimReduc}
    \end{figure}

    \subsubsection{Summary} 
    The overall procedures of FedVQCS are summarized in {\bf Procedure~\ref{alg:FedVQCS}}, while the local block compression and global block reconstruction processes are summarized in {\bf Procedure~\ref{alg:Compress}} and {\bf Procedure~\ref{alg:Reconstruction}}, respectively.
    In the remainder of this section, we elaborate more on the local block compression process (see Sec.~\ref{Sec:Comp}) and the global block reconstruction process (see Sec.~\ref{Sec:Reconst}). Note that FedVQCS can also be combined with other FL frameworks such as \cite{FL_R1,FL_R2,FL_R3} for further performance improvement.

    \subsection{Compression of Local Block Update}\label{Sec:Comp}
    The local block compression process of FedVQCS consists of three steps: (i) sparsification, (ii) dimensionality reduction, and (iii) vector quantization, as summarized in {\bf Procedure~\ref{alg:Compress}}.
    The details of each step performed by device $k$ for block $b$ in round $t$ are elaborated below.

    \subsubsection{Sparsification} In the sparsification step, each local block update is sparsified by nullifying all but the most significant $S_k$ entries in terms of their magnitudes, where $S_k<N$ is a sparsity level chosen by device $k$. 
    To compensate for the information loss during the sparsification, the nullified entries are added to the local block update in the next communication round, as done in \cite{Amiri:TSP} and \cite{Amiri:TWC}.
    With this compensation strategy, the local block update that needs to be sparsified at device $k$ is expressed as 
    \begin{align}\label{eq:true_local}
        \bar{\boldsymbol g}_k^{(t,b)} = {\boldsymbol g}_k^{(t,b)} + {\boldsymbol \Delta}_k^{(t-1,b)},
    \end{align}    
    where ${\boldsymbol \Delta}_k^{(t,b)} \in \mathbb{R}^{{N}}$ is a residual vector stored by device $k$ for block $b$ in round $t$.
    As a result, the local block update after the sparsification is denoted by $\tilde{\boldsymbol g}_{k}^{(t,b)} = {\sf Sparse}(\bar{\boldsymbol g}_{k}^{(t,b)})$ which has only $S_k$ nonzero entries. 
    Then, the residual vector is updated as 
	\begin{align}\label{eq:res}
        {\boldsymbol \Delta}_k^{(t,b)} = \bar{\boldsymbol g}_{k}^{(t,b)} - {\sf Sparse}(\bar{\boldsymbol g}_{k}^{(t,b)}).
    \end{align}
    Our sparsification step is summarized in Steps 1--3 of {\bf Procedure~\ref{alg:Compress}}. 



    \begin{algorithm}[t]
        \caption{Compression of Local Block Update}\label{alg:Compress}
        {\small
        {\begin{algorithmic}[1]
            \FUNC {${\sf BlkComp} \big({\boldsymbol g}_{k}^{(t,b)}, {\boldsymbol \Delta}_{k}^{(t-1,b)}\big)$}
                \STATE $\bar{\boldsymbol g}_{k}^{(t,b)} = {\boldsymbol g}_{k}^{(t,b)} + {\boldsymbol \Delta}_{k}^{(t-1,b)}$
                \STATE $\tilde{\boldsymbol g}_{k}^{(t,b)} = {\sf Sparse}(\bar{\boldsymbol g}_{k}^{(t,b)})$
                \STATE ${\boldsymbol \Delta}_{k}^{(t,b)} = \bar{\boldsymbol g}_{k}^{(t,b)}  - {\sf Sparse}( \bar{\boldsymbol g}_{k}^{(t,b)}  )$
                \STATE ${\boldsymbol x}_{k}^{(t,b)} = \alpha_{k}^{(t,b)} {\boldsymbol A}_{R_k}  \tilde{\boldsymbol g}_{k}^{(t,b)}$, where $\alpha_{k}^{(t,b)} = 1/\|\tilde{\boldsymbol g}_{k}^{(t,b)}\|$
                \STATE ${\boldsymbol x}_{k}^{(t,b)} = \big[({\boldsymbol v}_{k,1}^{(t,b)})^{\sf T},\ldots,({\boldsymbol v}_{k,P_k}^{(t,b)})^{\sf T}\big]^{\sf T}$
                \STATE $\hat{\boldsymbol v}_{k,p}^{(t,b)} = {\sf VQ}({\boldsymbol v}_{k,p}^{(t,b)})$,~$\forall p$
                \STATE $\Omega_{k}^{(t,b)} = {\sf Encode}\big(\{\hat{\boldsymbol v}_{k,p}^{(t,b)}\}_{p=1}^{P_k}, \alpha_{k}^{(t,b)}\big)$
                \STATE \textbf{return} $\Omega_{k}^{(t,b)}$ and ${\boldsymbol \Delta}_{k}^{(t,b)}$
            \ENDFUNC
        \end{algorithmic}}}
    \end{algorithm}
    \begin{algorithm}[t]
        \caption{Reconstruction of Global Block Update}\label{alg:Reconstruction}
        {\small
        {\begin{algorithmic}[1]
            \FUNC {$ {\sf BlkReconst}\big(\{\Omega_{k}^{(t,b)}\}_{k=1}^K\big)$}
                \STATE $\big(\{\hat{\boldsymbol v}_{k,p}^{(t,b)}\}_{p=1}^{P_k}, \alpha_k^{(t,b)} \big) = {\sf Decode}\big(\Omega_{k}^{(t,b)}\big)$, $\forall k$
                \STATE $\hat{\boldsymbol x}_k^{(t,b)} = \big[(\hat{\boldsymbol v}_{k,1}^{(t,b)})^{\sf T},\ldots,(\hat{\boldsymbol v}_{k,P_k}^{(t,b)})^{\sf T}\big]^{\sf T}$, $\forall k$
                \STATE ${\boldsymbol y}_{\mathcal{K}_g}^{(t,b)} = \sum_{k\in\mathcal{K}_g} \big({\rho_k}/{\alpha_{k}^{(t,b)}} \big) \hat{\boldsymbol x}_k^{(t,b)} $, $\forall g$
                \STATE $\hat{\boldsymbol g}_{\mathcal{K}_g}^{(t,b)} = {\sf SparseRecovery}\big({\boldsymbol A}_{R_g}, {\boldsymbol y}_{\mathcal{K}_g}^{(t,b)}\big)$, $\forall g$
                \STATE $\hat{\boldsymbol g}_{\mathcal{K}}^{(t,b)} = \sum_{g=1}^G \hat{\boldsymbol g}_{\mathcal{K}_g}^{(t,b)}$
                \STATE \textbf{return } $\hat{\boldsymbol g}_{\mathcal{K}}^{(t,b)}$
            \ENDFUNC
        \end{algorithmic}}}
    \end{algorithm}
    
    \subsubsection{Dimensionality reduction} 
    In the dimensionality reduction step, each local block update after the sparsification is projected onto a low-dimensional space as in CS \cite{Book:Math}. In other words, the dimension of the sparsified local block update $\tilde{\boldsymbol g}_k^{(t,b)}$ is reduced based on the CS theory, as illustrated in  Fig.~\ref{fig:DimReduc}.
    Let $R_k = \frac{N}{M_k} \geq 1$ be a dimensionality reduction ratio chosen by device $k$, where $M_k\leq N$ is the dimension after projection. Then, the compressed local block update ${\boldsymbol x}_{k}^{(t,b)}\in\mathbb{R}^{M_k}$ is obtained as
	\begin{align}\label{eq:compress_block}
        {\boldsymbol x}_{k}^{(t,b)}  = \alpha_{k}^{(t,b)} {\boldsymbol A}_{R_k} \tilde{\boldsymbol g}_{k}^{(t,b)},
    \end{align}
    where $\alpha_{k}^{(t,b)} \in \mathbb{R}$ is a scaling factor, and ${\boldsymbol A}_{R_k} \in \mathbb{R}^{M_k\times N}$ is a projection matrix.
    We set the scaling factor as $\alpha_{k}^{(t,b)}={1}/{\|\tilde{\boldsymbol g}_{k}^{(t,b)}\|}$ and the projection matrix as an IID Gaussian random matrix\footnote{{The design of the projection matrix can be optimized for further reducing the compression error. Other types of random matrices are also applicable in \eqref{eq:compress_block}.}} with zero mean and unit variance, i.e., $({\boldsymbol A}_{R_k} )_{m,n} \sim \mathcal{N}(0,1)$, $\forall m,n$.
     {Under these settings, the design of vector quantizer as well as the optimization of $S_k$ and $R_k$ will be elucidated in Sec.~\ref{Sec:FedVQCS+}.}
    We assume\footnote{In practice, this assumption can be realized by sharing the same $N\times N$ random projection matrix between the PS and the devices before the beginning of the training process. Then, in each local block update compression, device $k$ can determine its projection matrix by selecting the first $M_k$ rows of the shared projection matrix.} that the devices with the same dimensionality reduction ratio adopt the same projection matrix.
    Our dimensionality reduction step is summarized in Step 4 of {\bf Procedure~\ref{alg:Compress}}.

    \subsubsection{Vector quantization} 
    In the vector quantization step, each compressed local block update is quantized using a $Q_kM_k$-bit vector quantizer, where $Q_k$ is the number of quantization bits per entry chosen by device $k$.
    To avoid the computational complexity for quantizing a high-dimensional vector, the compressed local block update ${\boldsymbol x}_k^{(t,b)}$ is first partitioned into $P_k$ subvectors of dimension $L = \frac{M_k}{P_k}$, i.e., ${{\boldsymbol x}}_{k}^{(t,b)} = \big[({\boldsymbol v}_{k,1}^{(t,b)})^{\sf T},\ldots,({\boldsymbol v}_{k,P_k}^{(t,b)})^{\sf T}\big]^{\sf T}$, as in \cite{Du:20}.  
    Then, these subvectors are quantized in parallel using a $Q_k L$-bit vector quantizer.
    The resulting quantized subvector is given by
    \begin{align}\label{eq:quantize_block}
        \hat{\boldsymbol v}_{k,p}^{(t,b)} = {\sf Q}_{\mathcal{C}}({\boldsymbol v}_{k,p}^{(t,b)}),~p\in\{1,\ldots,P_k\},
    \end{align}
    where ${\sf Q}_{\mathcal{C}}:\mathbb{R}^L\rightarrow \mathcal{C}$ is a vector quantizer with a codebook $\mathcal{C}$ such that $|\mathcal{C}| \leq 2^{Q_kL}$.
    The optimal design of the vector quantizer will be discussed in Sec.~\ref{Sec:VQDesign}. 
    Our vector quantization step is summarized in Steps 5--6 of {\bf Procedure~\ref{alg:Compress}}. 
    After vector quantization, the quantized subvectors $\{\hat{\boldsymbol v}_{k,p}^{(t,b)}\}_{p=1}^{P_k}$ and the scaling factor $\alpha_k^{(t,b)}$ are encoded into digital bits, denoted by $\Omega_{k,b}^{(t)}$, for digital communications.

    \subsection{Reconstruction of Global Block Update}\label{Sec:Reconst}
    The global block reconstruction process of FedVQCS consists of two steps: (i) group-wise aggregation, and (ii) sparse signal recovery, as summarized in {\bf Procedure~\ref{alg:Reconstruction}}.
    The details of each step performed by the PS for block $b$ in round $t$ are elaborated below.
    
    \subsubsection{Group-wise aggregation}
    In the group-wise aggregation step, 
    the PS groups the compressed local block updates with the same dimension and, then, aggregates only the compressed updates in the same group.
    Meanwhile, to facilitate an accurate reconstruction of the aggregated block update, the PS limits the number of the compressed updates in each group to be less than or equal to $K^\prime$. 
    By adjusting $K^\prime$, the PS can control the tradeoff between the complexity and accuracy of the global block reconstruction, as will be justified later.

    Let $\mathcal{K}_g$ be the indices of the compressed local block updates in group $g\in\{1,\ldots,G\}$, where $G$ is the number of groups, and $\mathcal{K}_1,\ldots,\mathcal{K}_G$ are mutually exclusive subsets of $\mathcal{K}$ such that $\mathcal{K} = \bigcup_{g=1}^G \mathcal{K}_g$.
    For ease of exposition, we assume that $|\mathcal{K}_g| = K^\prime$, $\forall g$. 
    Under error-free reception of the transmitted data $\{\Omega_{k}^{(t,b)}\}_{k\in\mathcal{K}}$, the PS obtains the quantized subvectors $\{\hat{\boldsymbol v}_{k,p}^{(t,b)}\}_{p=1}^{P_k}$ and the scaling factor $\alpha_k^{(t,b)}$ by decoding $\Omega_{k}^{(t,b)}$. 
    Let  ${\boldsymbol d}_{k,p}^{(t,b)} \triangleq \hat{\boldsymbol v}_{k,p}^{(t,b)} - {\boldsymbol v}_{k,p}^{(t,b)} $ be a quantization error vector. 
    Then, from $\hat{\boldsymbol v}_{k,p}^{(t,b)} = {\boldsymbol v}_{k,p}^{(t,b)} + {\boldsymbol d}_{k,p}^{(t,b)}$, the compressed local block update $\hat{\boldsymbol x}_{k}^{(t,b)}$ is expressed as
    \begin{align}
        \hat{\boldsymbol x}_{k}^{(t,b)} 
        &= \big[(\hat{\boldsymbol v}_{k,1}^{(t,b)})^{\sf T},\ldots,(\hat{\boldsymbol v}_{k,P}^{(t,b)})^{\sf T}\big]^{\sf T} 
        = {\boldsymbol x}_{k}^{(t,b)} + {\boldsymbol d}_{k}^{(t,b)} \nonumber \\&= \alpha_{k}^{(t,b)} {\boldsymbol A}_{R_k} \tilde{\boldsymbol g}_{k}^{(t,b)} + {\boldsymbol d}_{k}^{(t,b)},        \label{eq:x_distortion}
    \end{align}   
    where ${\boldsymbol d}_k^{(t,b)} = \big[({\boldsymbol d}_{k,1}^{(t,b)})^{\sf T},\ldots,({\boldsymbol d}_{k,P}^{(t,b)})^{\sf T}\big]^{\sf T}$. 
    Because $R_k=R_g$, $\forall k\in\mathcal{K}_g$, the PS can aggregate the compressed local block updates in group $g$ as 
    \begin{align}
        {\boldsymbol y}_{\mathcal{K}_g}^{(t,b)} 
        = \sum_{k\in\mathcal{K}_g} \frac{\rho_k}{\alpha_{k}^{(t,b)}} \hat{\boldsymbol x}_{k}^{(t,b)} 
        = {\boldsymbol A}_{R_g}\tilde{\boldsymbol g}_{\mathcal{K}_g}^{(t,b)}  + {\boldsymbol d}_{\mathcal{K}_g}^{(t,b)}, \label{eq:aggregate_FedVQCS}
    \end{align}
    where $\tilde{\boldsymbol g}_{\mathcal{K}_g}^{(t,b)} = \sum_{k\in\mathcal{K}_g} \rho_k \tilde{\boldsymbol g}_{k}^{(t,b)}$ is the aggregated block update of group $g$ for block $b$ in round $t$ and ${\boldsymbol d}_{\mathcal{K}_g}^{(t,b)} = \sum_{k\in\mathcal{K}_g} ({\rho_k}/{\alpha_{k}^{(t,b)}}){\boldsymbol d}_{k}^{(t,b)}$. 

    \subsubsection{Sparse signal recovery}
    In the sparse signal recovery step, the PS estimates the aggregated block update $\tilde{\boldsymbol g}_{\mathcal{K}_g}^{(t,b)}$ from its noisy linear observation ${\boldsymbol y}_{\mathcal{K}_g}^{(t,b)}$ in \eqref{eq:aggregate_FedVQCS} for every group $g$.
    As can be seen in  \eqref{eq:aggregate_FedVQCS}, the estimation of $\tilde{\boldsymbol g}_{\mathcal{K}_g}^{(t,b)}$ from ${\boldsymbol y}_{\mathcal{K}_g}^{(t,b)}$ is a well-known sparse signal recovery problem \cite{Book:Math}.
    Motivated by this fact, the PS employs a sparse signal recovery algorithm (e.g., \cite{EMGAMP,IHT}) to estimate $\tilde{\boldsymbol g}_{\mathcal{K}_g}^{(t,b)}$ from ${\boldsymbol y}_{\mathcal{K}_g}^{(t,b)}$ for every group $g$. 
    The sparse signal recovery algorithm considered in our work will be elucidated in Sec.~\ref{Sec:Simul}. 
    If the sparsity level of  $\tilde{\boldsymbol g}_{\mathcal{K}_g}^{(t,b)}$ is sufficiently smaller than the dimension of ${\boldsymbol y}_{\mathcal{K}_g}^{(t,b)}$, which will be justified in Sec.~\ref{Sec:Param},
    the PS is able to attain an accurate estimate of $\tilde{\boldsymbol g}_{\mathcal{K}_g}^{(t,b)}$.
    After the recovery of the aggregated block updates for all groups, the PS reconstructs the global block update as $\hat{\boldsymbol g}_{\mathcal{K}}^{(t,b)} = \sum_{g=1}^G \hat{\boldsymbol g}_{\mathcal{K}_g}^{(t,b)}$.
    Finally, the PS obtains the global model update $\hat{\boldsymbol g}_{\mathcal{K}}^{(t)}$ by concatenating all the global block updates $\{\hat{\boldsymbol g}_{\mathcal{K}}^{(t,b)}\}_{b=1}^B$ (see Step 14 of {\bf Procedure~\ref{alg:FedVQCS}}).

    \vspace{1mm}
    {\bf Remark 1 (Communication overhead of FedVQCS):}
    In FedVQCS, each device $k$ transmits the digital bits representing the quantized subvectors $\{\hat{\boldsymbol v}_{k,p}^{(t,b)}\}_{p=1}^{P_k}$ and the scaling factor $\alpha_k^{(t,b)}$ for every block $b\in\{1,\ldots,B\}$ in round $t$. 
    Because every quantized subvector is represented by $Q_k L$ bits, the total number of digital bits required for transmitting the local model update at each device is given by $(Q_k LP_k + 32)B = (Q_k M_k + 32)B$ for every round. 
    If $Q_k M_k \gg 32$, the communication overhead for device $k$ is $\frac{Q_k M_k B}{NB} = \frac{Q_k}{R_k}$ bits per local model entry.
    Owing to this feature, FedVQCS allows each device to not only reduce the communication overhead for transmitting its local model update but also control this overhead by adjusting the values of $Q_k$ and $R_k$ based on the capacity of the wireless link. 
    Therefore, FedVQCS is a powerful framework for enabling wireless FL with limited link capacity.  
    The optimization of $Q_k$ and $R_k$ for a given link capacity will be elucidated in Sec.~\ref{Sec:Param}.
     {It should also be noted that in FedVQCS, the PS and devices need to share some information such as $\mathcal{C}$, ${\boldsymbol A}_{R_k}$, and $B$ before the beginning of training process. Fortunately, the communication overhead required for sharing this information is marginal compared to that required for transmitting local model updates during the training process. Hence, in this work, we consider only the significant communication overhead that occurs during the local model update transmission in FL.} 
        
    \vspace{1mm}
    {\bf Remark 2 (Comparison with existing FL frameworks):} 
    The existing FL frameworks in \cite{UVeqFed,UVeqFed2,Du:20} consider vector quantization for compressing the local model updates, but without applying any dimensionality reduction technique.
    Unlike these frameworks, FedVQCS applies the CS-based dimensionality reduction before the vector quantization, which reduces the dimension of the local model update by allowing the transmission of its most significant entries only.
    In \cite{Aji:17,Wangni:18,Lin:18,Amiri:TSP,Amiri:TWC}, it has been reported that the accuracy of FL does not severely degrade when sending a sufficient number of the significant entries.
    Therefore, FedVQCS achieves further compression of the local model updates compared with the frameworks in \cite{UVeqFed,UVeqFed2,Du:20}, without significantly sacrificing the learning accuracy. 
    The existing FL frameworks in \cite{QCS1,QCS2,FedQCS} employ CS-based dimensionality reduction but rely solely on scalar quantization. In contrast, FedVQCS empolys vector quantization, which generally outperforms scalar quantization in terms of quantization error. Furthermore, FedVQCS optimizes key parameters such as the number of quantization bits per entry $Q$, a sparsity level $S$, and a dimensionality reduction ratio $R$ to minimize the reconstruction error at the PS (see Sec.~\ref{Sec:Param}). This optimization approach has not been thoroughly explored in the FL literature. Therefore, FedVQCS enables a more accurate reconstruction of the global model update compared to the QCS-based FL frameworks in \cite{QCS1,QCS2,FedQCS}.

    \section{Optimization and Convergence Analysis of FedVQCS}\label{Sec:FedVQCS+}
    Here, we first optimize the design of a vector quantizer for the local block compression process of FedVQCS and then optimize the key design parameters of this compression process for minimizing the reconstruction error of the global model update at the PS. 
    We also characterize the convergence rate of FedVQCS, while considering the reconstruction error at the PS.

    \subsection{Optimal Design of Vector Quantizer}\label{Sec:VQDesign}
    In FedVQCS, a proper design of the vector quantizer in \eqref{eq:quantize_block} is critical for reducing the reconstruction error of the global model update at the PS. 
    The underlying challenge of the vector quantizer design is that the exact distribution of the quantizer input depends on many factors such as the global model choice, the local training data distribution, and the loss function type. 
    Moreover, the optimal vector quantizer may differ across partitions, blocks, devices, and communication rounds, which can lead to additional communication overhead for transmitting the information of the quantizer between the PS and the devices.

    Fortunately, the above challenges can be readily addressed in FedVQCS.
    First of all, it is reported in \cite{FedQCS} that a sparsified local block update $\tilde{\boldsymbol g}_{k}^{(t,b)}$ can be modeled as an IID random vector whose entry follows a Bernoulli Gaussian-mixture distribution. 
    Hence, for a large $N$, the projected local model update ${\boldsymbol x}_{k}^{(t,b)} = \alpha_{k}^{(t,b)} {\boldsymbol A}_{R_k} \tilde{\boldsymbol g}_{k}^{(t,b)}$ can be modeled as an $M_k$-dimensional IID Gaussian random vector with zero mean and unit variance by the central limit theorem, provided that ${\boldsymbol A}_{R_k}$ is an IID Gaussian random matrix with $\alpha_{k}^{(t,b)}={1}/{\|\tilde{\boldsymbol g}_{k}^{(t,b)}\|}$. 
    This implies that every subvector of ${\boldsymbol x}_{k}^{(t,b)}$ can also be modeled as an $L$-dimensional IID Gaussian random vector whose distribution is same for all partitions, blocks, devices, and communication rounds, i.e., ${\boldsymbol v}_{k,p}^{(t,b)}\sim\mathcal{N}({\boldsymbol 0}_L,{\boldsymbol I}_L)$, $\forall k,p,t,b$. 
    Motivated by this observation, in FedVQCS, we use the vector quantizer optimized for the distribution of ${\boldsymbol v}\sim\mathcal{N}({\boldsymbol 0}_L,{\boldsymbol I}_L)$ for a given number of quantization bits. 
    Owing to this feature, both the PS and devices in FedVQCS can utilize the same optimal quantizer by sharing the number of quantization bits.

    \subsubsection{Quantizer design problem}
    To determine the optimal vector quantizer for the  distribution of ${\boldsymbol v}\sim\mathcal{N}({\boldsymbol 0}_L,{\boldsymbol I}_L)$, we consider a {\em shape-gain} quantizer, which is effective in quantizing an IID Gaussian random vector while enabling the efficient construction of the optimal codebook \cite{GainShape}.
    When employing this quantizer, the shape of ${\boldsymbol v}$, defined as ${\boldsymbol s}={\boldsymbol v}/\|{\boldsymbol v}\|$, and its gain defined as $h = \|{\boldsymbol v}\|$  are independently quantized.
    Let ${\sf Q}_{\mathcal{C}_{\rm s}} ( {\boldsymbol s})=  \underset{\hat{\boldsymbol s} \in \mathcal{C}_{\rm s}}{\arg\!\min}~ \| {\boldsymbol s} - \hat{\boldsymbol s} \|^2 $ and  ${\sf Q}_{\mathcal{C}_{\rm h}} (h) = \underset{\hat{h}\in\mathcal{C}_{\rm h}}{\arg\!\min}~ | h - \hat{h}|^2$ be the shape and gain quantizers using the minimum Euclidean distance criterion, respectively.
    Then, the output of the shape-gain quantizer will be $\hat{\boldsymbol v} = \hat{h}\hat{\boldsymbol s}$, where $\hat{\boldsymbol s} = {\sf Q}_{\mathcal{C}_{\rm s}}({\boldsymbol s})$ and $\hat{h} = {\sf Q}_{\mathcal{C}_{\rm h}}(h)$.
    As such, the MSE of the shape-gain quantizer can be approximated by \cite{GainShape}
    \begin{align}\label{eq:GainShapeMSE}
        \mathbb{E}\big[\| {\boldsymbol v} - \hat{h}\hat{\boldsymbol s}\|^2\big] 
        &\approx 
        L \mathbb{E}\big[\| {\boldsymbol s} - \hat{\boldsymbol s} \|^2\big]
        + \mathbb{E}\big[| {h} - \hat{h}|^2\big] \nonumber \\
        &= L\cdot {\sf MSE}({\boldsymbol s};\mathcal{C}_{\rm s}) + {\sf MSE}(h;\mathcal{C}_{\rm h}),
    \end{align}
    where ${\sf MSE}(h;\mathcal{C}_{\rm h})  =  \mathbb{E}\big[| h - {\sf Q}_{\mathcal{C}_{\rm h}}(h)|^2\big]$ and   ${\sf MSE}({\boldsymbol s};\mathcal{C}_{\rm s}) = \mathbb{E}\big[ \| {\boldsymbol s} - {\sf Q}_{\mathcal{C}_{\rm s}}({\boldsymbol s}) \|^2\big]$.
    By utilizing this approximation, we formulate the shape-gain quantizer design problem for a given number of quantization bits $QL$ as follows:
    \begin{align}\label{eq:VQ_design1}
        &\underset{{\mathcal{C}_{\rm s}},{\mathcal{C}_{\rm h}}}{\arg\!\min}~
        L\cdot {\sf MSE}({\boldsymbol v}/\|{\boldsymbol v}\|;\mathcal{C}_{\rm s}) + {\sf MSE}(\|{\boldsymbol v}\|;\mathcal{C}_{\rm h}), \nonumber \\
        &\text{~~~s.t.~~~} {\boldsymbol v}\sim\mathcal{N}({\boldsymbol 0}_L,{\boldsymbol I}_L), ~|\mathcal{C}_{\rm s}| \leq 2^{Q_{\rm s} },~|\mathcal{C}_{\rm h}| \leq 2^{Q_{\rm h} },\nonumber \\ &~~~~~~~~\,~Q_{\rm s} + Q_{\rm h} \leq QL.
    \end{align}         
    This problem is solved using the following strategy: First, we separately determine the MSE-optimal shape and gain quantizers for a fixed bit allocation $(Q_{\rm s},Q_{\rm h})$ and, then, we derive the optimal bit allocation $(Q_{\rm s}^\star ,Q_{\rm h}^\star )$ for the optimal shape and gain quantizers.

    \subsubsection{Design of shape quantizer}
    When the number of shape quantization bits is given by $Q_{\rm s}$, the MSE-optimal  shape quantizer  is obtained by solving the following problem:
    \begin{align}\label{eq:SQ_design}
        \mathcal{C}_{\rm s}^\star = \underset{|\mathcal{C}_{\rm s}| \leq 2^{Q_{\rm s}}} {\arg\!\min} ~{\sf MSE}({\boldsymbol s};\mathcal{C}_{\rm s}),
    \end{align} 
    where ${\boldsymbol s} = {\boldsymbol v}/\|{\boldsymbol v}\|$ with ${\boldsymbol v}\sim\mathcal{N}({\boldsymbol 0}_L,{\boldsymbol I}_L)$.
    Unfortunately, it is hard to characterize the exact solution of this problem; thus, as an alternative, we consider an approximate solution by replacing the Euclidean distance with the squared chordal distance, as in \cite{Du:20}.
    Then the approximate solution of \eqref{eq:SQ_design} is obtained as an even Grassmannian codebook $\mathcal{C}_{\rm s}^\star$ that can be constructed as follows \cite{Du:20}: 
    \begin{itemize}
        \item Solve the Grassmannian line packing problem formulated as \cite{Grassline}:
        \begin{align}\label{eq:Grassline}
            \max_{\mathcal{C}_{\rm s}^{+}}  \min_{ \hat{\boldsymbol s}\neq\hat{\boldsymbol s}^{\prime},\hat{\boldsymbol s},\hat{\boldsymbol s}^{\prime}\in\mathcal{C}_{\rm s}^{+}}d(\hat{\boldsymbol s},\hat{\boldsymbol s}^{\prime}),~~\text{s.t.}~~|\mathcal{C}_{\rm s}^{+}| = 2^{Q_{\rm s}-1},
        \end{align}
        where 
        $d(\hat{\boldsymbol s}, \hat{\boldsymbol s}^{\prime}) = \sqrt{1-|\hat{\boldsymbol s}^{\sf T}\hat{\boldsymbol s}^{\prime}|^2}$.

        \item Construct the even Grassmannian codebook as $\mathcal{C}_{\rm s}^\star = \mathcal{C}_{\rm s}^{+}\bigcup\mathcal{C}_{\rm s}^{-}$ where $\mathcal{C}_{\rm s}^{-} = \{{\boldsymbol s}: -{\boldsymbol s}\in\mathcal{C}_{\rm s}^{+} \}$.
    \end{itemize}
    We determine the optimal shape quantizer by using the even Grassmannian codebook $\mathcal{C}_{\rm s}^\star$ constructed above. 
    The MSE of the above shape quantizer is characterized, as given in the following lemma:
    \begin{lem}\label{lem:MSE_shape}
        For large $L$, the MSE of the even Grassmannian codebook $\mathcal{C}_{\rm s}^\star$ with size $2^{Q_{\rm s}}$ is bounded as
        \begin{align}\label{eq:MSE_shape}
            2 - 2\sqrt{1-\frac{L-1}{L+1}2^{-\frac{2(Q_{\rm s}-1)}{L-1}}}
            \leq {\sf MSE}({\boldsymbol s};\mathcal{C}_{\rm s}^\star)
            \leq 2^{-\frac{2(Q_{\rm s}-1)}{L-1} + 1}.
        \end{align}
    \end{lem}
    \begin{IEEEproof}
        See Appendix~\ref{Apdx:MSE_shape}.
    \end{IEEEproof}

    \subsubsection{Design of gain quantizer}
    When the number of gain quantization bits is given by $Q_{\rm h}$, the MSE-optimal gain quantizer is obtained by solving the following problem:
    \begin{align}\label{eq:GQ_design}
        \mathcal{C}_{\rm h}^\star = \underset{|\mathcal{C}_{\rm h}| \leq 2^{Q_{\rm h}}} {\arg\!\min} ~{\sf MSE}(h;\mathcal{C}_{\rm h}),
    \end{align} 
    where $h = \|{\boldsymbol v}\|$. From ${\boldsymbol v}\sim\mathcal{N}({\boldsymbol 0}_L,{\boldsymbol I}_L)$, the probability density function of the gain $h = \|{\boldsymbol v}\|$ is given by
    \begin{align}\label{eq:gain_pdf}
        f_{h}(z) = \frac{2z^{L-1}\exp{(\frac{-z^2}{2})}}{\Gamma(L/2)2^{L/2}},
    \end{align}
    where $\Gamma(r) = \int_{0}^{\infty}t^{r-1}e^{-t} {\rm d}t$.
    Because this distribution is strictly log-concave, the Lloyd--Max algorithm  in \cite{LloydMax} converges to a globally optimal quantizer in terms of minimizing the MSE \cite{Lloyd1}.
    Hence, when $Q_{\rm h} > 0 $, we determine the optimal gain quantizerby applying the Lloyd--Max algorithm for the distribution in \eqref{eq:gain_pdf}.
    When $Q_{\rm h}=0$, our strategy is to approximate the gain $h=\|{\boldsymbol v}\|$ as its expected value $\mathbb{E}[h] = \sqrt{2}\frac{\Gamma\left((L+1)/2\right)}{\Gamma\left(L/2\right)}$ based on its distribution.
    Considering these two cases, the MSE of the above gain quantizer is characterized, as given in the following lemma:
    \begin{lem}\label{lem:MSE_gain}
        The MSE of the optimal gain quantizer is approximated by
        \begin{align}\label{eq:MSE_gain}
            {\sf MSE}(h;\mathcal{C}_{\rm h}^\star)
            \approx \begin{cases}
                \chi_L 2^{-2(Q_{\rm h}+1)}, & Q_{\rm h}> 0, \\
                L - \frac{2\pi}{\beta^2\left(\frac{L}{2},\frac{1}{2}\right)}, & Q_{\rm h}=0,
            \end{cases}
      \end{align}
      where $\chi_L = \frac{3^{\frac{L}{2}}\Gamma^3(\frac{L+2}{6})}{2\Gamma(\frac{L}{2})}$.
    \end{lem}
    \begin{IEEEproof}
        The MSE of the Lloyd--Max codebook with size $Q_{\rm h} \gg 1$, namely $\mathcal{C}_{\rm h}^\star$, is approximated by \cite{MSELloyd}
        \begin{align}
            {\sf MSE}(h;\mathcal{C}_{\rm h}^\star)
            &\approx \frac{1}{12}\left(\int_{0}^{\infty}f_h(z){\rm d}z\right)^32^{-2Q_{\rm h}} \nonumber \\&= \frac{3^{\frac{L}{2}}\Gamma^3\left(\frac{L + 2}{6}\right)}{8\Gamma\left(\frac{L}{2}\right)}2^{-2Q_{\rm h}}. \label{eq:MSE_gain_LM}
        \end{align}
        We use the above expression for approximating the MSE of the Lloyd--Max quantizer for $Q_{\rm h}>0$. The MSE of $\hat{h} = \mathbb{E}[h]$ is computed as
        \begin{align}
            \mathbb{E}[(h-\mathbb{E}[h])^2] = \mathbb{E}[h^2] - (\mathbb{E}[h])^2 = L - \frac{2\pi}{\beta^2(\frac{L}{2},\frac{1}{2})}, \label{eq:MSE_gain_mean}
        \end{align}
        where $\beta(r,s) = \int_0^1 t^{r-1}(1-t)^{s-1}{\rm d}t$. This expression gives the MSE of the optimal gain quantizer for $Q_{\rm h}=0$.
    \end{IEEEproof}

    \subsubsection{Optimal bit allocation}
    Now, we derive the optimal bit allocation for the optimal shape and gain quantizers based on the MSE characterizations in \eqref{eq:MSE_shape} and \eqref{eq:MSE_gain}. 
    This result is given in the following theorem:
    
    \vspace{1mm}
    \begin{thm}\label{Thm:Bit_opt}
        If both the upper bound in \eqref{eq:MSE_shape} and the approximation in \eqref{eq:MSE_gain} hold with equality, the optimal bit allocation $\big(Q_{\rm s}^\star , Q_{\rm h}^{\star} \big)$ in \eqref{eq:VQ_design1} is given by
        \begin{align} \label{eq:Bit_Opt}
            &\big(Q_{\rm s}^\star , Q_{\rm h}^{\star} \big) = 
            \begin{cases}
                \big(QL-H_{L,Q}, H_{L,Q}\big), &
                \text{if}~F_{L,Q} \leq 0,
                \\
                \big(QL,0\big), & \text{if}~F_{L,Q} >  0,
                \\
            \end{cases}
        \end{align}
        where $H_{L,Q}= \frac{L-1}{2L} \log_{2} \left(\frac{L-1}{2L} \chi_L\right) + Q - 1$ and
        \begin{align} 
            F_{L,Q} &= L2^{-\frac{2(QL-1)}{L-1} + 1}  \big( 2^{\frac{2H_{L,Q} }{L-1} + 1} -1\big)  \nonumber \\&~ ~~+ \chi_L 2^{-2(H_{L,Q} +1)} - L+\frac{2\pi}{\beta^2\left(\frac{L}{2}, \frac{1}{2}\right)}.
        \end{align}
    \end{thm}
    \begin{IEEEproof}
    From {Lemma 1} and {Lemma 2}, the bit allocation problem with $Q_{\rm h}>0$ is 
    \begin{align}\label{eq:MSE1}
        ({\bf P}_1)~~&\underset{Q_{\rm s} , Q_{\rm h}}{\arg\!\min}
        ~L2^{-\frac{2(Q_{\rm s}-1)}{L-1} + 1} + \chi_L 2^{-2(Q_{\rm h}+1)},  \nonumber \\
        &\text{~~~s.t.~~~}  Q_{\rm s} + Q_{\rm h} \leq QL, ~Q_{\rm h}>0.
    \end{align}
    Because $({\bf P}_1)$ is convex \cite{Du:20}, the optimal solution is readily derived by leveraging Karush-Kuhn-Tucker (KKT) conditions:
    \begin{align}
        \lambda^{\star} + \frac{\partial F_{L,1}(Q_{\rm s},Q_{\rm h})}{\partial Q_{\rm s}} = 0, ~~ 
        \lambda^{\star} + \frac{\partial F_{L,1}(Q_{\rm s},Q_{\rm h})}{\partial Q_{\rm h}} = 0,
    \end{align}
    where $F_{L,1}(Q_{\rm s},Q_{\rm h})$ is the objective function of $({\bf P}_1)$, and $\lambda^{\star}$ is a Lagrange multiplier.
    From the KKT conditions, the solution of $({\bf P}_1)$ is given by $\big(Q_{\rm s}, Q_{\rm h}\big) =  ( QL-H_{L,Q}, H_{L,Q})$, where 
    \begin{align} 
        H_{L,Q}= \frac{L-1}{2L} \log_{2} \left(\frac{L-1}{2L} \chi_L\right) + Q - 1.
        \label{eq:MSE1_sol} 
    \end{align}
    From {Lemma 1} and {Lemma 2}, the bit allocation problem with $Q_{\rm h}=0$ is formulated as
    \begin{align}\label{eq:MSE2}
        ({\bf P}_2)~~&\underset{Q_{\rm s} \leq Q L}{\arg\!\min}
        ~L2^{-\frac{2(Q_{\rm s}-1)}{L-1} + 1} +    L-\frac{2\pi}{\beta^2\left(\frac{L}{2}, \frac{1}{2}\right)}.
    \end{align}
    Because the objective function of $({\bf P}_2)$ is a decreasing function of $Q_{\rm s}$, the solution is given by $Q_{\rm s} = QL$.
    The optimal bit allocation in \eqref{eq:Bit_Opt} can be obtained by comparing $F_{L,1}(QL-H_{L,Q}, H_{L,Q} )$ and $F_{L,2}({Q}L,0)$.  This completes the proof.
    \end{IEEEproof}
    \vspace{1mm}

    From {Theorem 1}, the MSE of the optimal shape-gain quantizer adopted in FedVQCS is approximated by
        \begin{align}\label{eq:MSE_VQ}
          \sigma_{L,Q}^2 \approx \begin{cases}
            L2^{-\frac{2(QL - H_{L,Q}-1)}{L-1} + 1} + \chi_L 2^{-2(H_{L,Q} +1)},\!\!\!\!\!&F_{L,Q} \leq 0, \\
            L 2^{-\frac{2(Q L-1)}{L-1} + 1} +  L - \frac{2\pi}{ \beta^2\left(\frac{L}{2}, \frac{1}{2}\right)},&F_{L,Q} > 0.\\
        \end{cases}
    \end{align}
    The MSE in \eqref{eq:MSE_VQ} clearly demonstrates that the increase in the number of quantization bits reduces the vector quantization error during the local block compression.
    Therefore, any increase in the capacity of wireless links will lead to a reduction in the reconstruction error of the global block update at the PS in FedVQCS.

    \subsection{Parameter Optimization for Local Block Compression}\label{Sec:Param}
    The local block compression process of FedVQCS has three design parameters: (i) the number of quantization bits per entry, $Q$, (ii) a sparsity level, $S$, and (iii) a dimensionality reduction ratio, $R$.
    These parameters are closely related to the wireless link capacity as well as the reconstruction error of the global model update at the PS. 
    Motivated by this fact, we optimize the above parameters for minimizing the reconstruction error at the PS under the constraint of the wireless link capacity. 
    In this optimization, we assume that the maximum number of digital bits that can be reliably transmitted from device $k$ to the PS in each communication round is given by $C_k\bar{N} > 0$.
     {Here, $C_k$ is determined solely by the wireless link capacity, which can be interpreted as a target compression ratio. For example, if $C_k$ is given by $0.5$ bits per entry according to the wireless link capacity, the $\bar{N}$-dimensional local model update needs to be compressed, so that it can be transmitted using only $\bar{N}/2$ bits in each iteration.}
    We also assume that $C_k$ is known at both the PS and device $k$.
    Our strategy for parameter optimization is as follows: We first characterize the best choice of $Q_k^\star(R_k)$ and $S_k^\star (R_k)$ as a function of $R_k$. We then determine the best dimensionality reduction ratio $R_k^\star$ when $Q_k = Q_k^\star(R_k)$ and $S_k = S_k^\star(R_k)$.
    In the remainder of this subsection, we omit the indices $t$ and $b$ for ease of exposition. 

    \subsubsection{Characterization of $Q_k^\star (R_k)$ and $S_k^\star (R_k)$} 
    As discussed in {Remark 1}, the communication overhead of FedVQCS for device $k$ is $\frac{Q_k}{R_k}\bar{N}$ bits.
    This overhead needs to be less than or equal to $C_k\bar{N}$ for error-free transmission of the compressed local block updates from device $k$ to the PS.  
    Therefore, the optimal number of quantization bits to minimize the quantization error for a fixed $R_k$ is 
    \begin{align}\label{eq:Q}
         Q_{k}^\star (R_k) = C_k R_k.
    \end{align} 
    To determine the proper sparsity level, we consider the fact that the PS estimates the aggregated block update $\tilde{\boldsymbol g}_{\mathcal{K}_g}= \sum_{k\in\mathcal{K}_g} \rho_k\tilde{\boldsymbol g}_{k}$ from its noisy linear observation ${\boldsymbol y}_{\mathcal{K}_g} = {\boldsymbol A}_{R_g}\tilde{\boldsymbol g}_{\mathcal{K}_g} + {\boldsymbol d}_{\mathcal{K}_g}$ during the global block reconstruction, where $({\boldsymbol A}_{R_g})_{m,n}\sim\mathcal{N}(0,1)$.
    Let $\tilde{S}_g$ be the sparsity level of the aggregated block update $\tilde{\boldsymbol g}_{\mathcal{K}_g}$.
    According to the CS theory \cite{Book:Math}, a well-established sparse signal recovery algorithm can find a unique solution of the above sparse signal recovery problem with high probability, if
    \begin{align}\label{eq:S_}
        R_k < \frac{N}{2\tilde{S}_g\log(N/\tilde{S}_g)},
    \end{align}
    for a large $N$, moderately large $\tilde{S}_g$, and large $N/\tilde{S}_g$.
    Inspired by this result, we shall use the inequality \eqref{eq:S_} to check the feasibility of the accurate reconstruction of $\tilde{\boldsymbol g}_{\mathcal{K}_g}$ from ${\boldsymbol y}_{\mathcal{K}_g}$.
    Recall that during the global block reconstruction process of FedVQCS, the PS groups the local model updates with the same dimensionality reduction ratio (i.e., $R_k = R_{g}$, $\forall k\in\mathcal{K}_g$), as explained in Sec.~\ref{Sec:Reconst}.
    Hence, the local block updates that belong to the same group have the same sparsity level  (i.e., $S_{k}^\star(R_k) = S_{g}^\star$, $\forall k\in\mathcal{K}_g$).
    From this fact, one can easily show that the condition in \eqref{eq:S_} always holds for all possible $\tilde{\boldsymbol g}_{\mathcal{K}_g}$ if $R_k < \frac{N}{{2 K^\prime S_k\log(N/(K^\prime S_k))}}$.
    Therefore, we determine the best sparsity level for a fixed $R_k$ as
    \begin{align}
        S_{k}^\star(R_k) = \underset{S}{\arg\!\max}~S,~~\text{s.t.}~~ R_k < \tfrac{N}{{2 K^\prime S \log(N/(K^\prime S))}}. \label{eq:S_2}
    \end{align}

    \subsubsection{Optimization of $R_k$} 
    To facilitate the group-wise aggregation strategy explained in Sec.~\ref{Sec:Reconst}, the dimensionality reduction ratio is assumed to be chosen among a candidate set $\mathcal{R}^{\rm c} = \{R_1^{\rm c},\ldots,$ $R_C^{\rm c}: R_i^{\rm c}\geq 1, \forall i\}$, shared by all the devices.
    Under this assumption, we aim to find the optimal choice of $R_k \in \mathcal{R}^{\rm c}$ that can minimize the MSE of the global block reconstruction at the PS.
    This optimization problem can be formulated as 
    \begin{gather}
        R_k^\star = \argmin_{R_k \in \mathcal{R}^{\rm c}} \mathbb{E}\left[\|\bar{\boldsymbol g}_{\mathcal{K}_g} - \hat{\boldsymbol g}_{\mathcal{K}_g}\|^2\right], \label{eq:Prob_SRQ}
    \end{gather}
    where $\bar{\boldsymbol g}_{\mathcal{K}_g}= \sum_{k\in\mathcal{K}_g} \rho_k \bar{\boldsymbol g}_k$, and the expectation is taken with respect to vector quantization error.
    It is generally challenging for the devices to locally solve the problem in \eqref{eq:Prob_SRQ} because of a lack of knowledge of the local block updates sent by other devices in the same group. In addition, the estimate of the aggregated block update, $\hat{\boldsymbol g}_{\mathcal{K}_g}$, depends on the sparse signal recovery algorithm chosen at the PS. 
    To overcome these challenges, we consider the ideal scenario in which the PS perfectly finds the support set\footnote{The support set of a vector ${\boldsymbol A}$ is defined as $\mathcal{T}({\boldsymbol A}) = \{i:a_i\neq 0\}$, which is the index set of the nonzero entries in  ${\boldsymbol A}$.} of $\tilde{\boldsymbol g}_{\mathcal{K}_g}$ by employing a powerful CS recovery algorithm, and, then, it minimizes the squared error $\|\tilde{\boldsymbol g}_{\mathcal{K}_g} - \hat{\boldsymbol g}_{\mathcal{K}_g}\|^2$ by solving the sparse signal recovery problem in \eqref{eq:aggregate_FedVQCS}.
    Under this consideration, we characterize the upper bound of the objective function in \eqref{eq:Prob_SRQ}, as given in the following theorem:
    \vspace{1mm}
    \begin{thm}\label{Thm:MSE_upper}
        Let $\hat{\boldsymbol g}_{\mathcal{K}_g}$ be the least squares solution of the sparse recovery problem in \eqref{eq:aggregate_FedVQCS} obtained with the knowledge of the support set of $\tilde{\boldsymbol g}_{\mathcal{K}_g}$. If $\mathbb{E}[\|{\boldsymbol d}_{k}\|^2] = \sigma_{L,Q_k}^2$, $\forall k\in\mathcal{K}_g$, for a large $\frac{N}{R_k} \gg 1$ and $\tilde{S}_g \gg 1$, the MSE of $\hat{\boldsymbol g}_{\mathcal{K}_g}$ is upper bounded as follows:
        \begin{align}\label{eq:MSE_upper}
            &\mathbb{E}[\|\bar{\boldsymbol g}_{\mathcal{K}_g} - \hat{\boldsymbol g}_{\mathcal{K}_g}\|^2]  \nonumber \\
            &\leq 
            K^\prime   \sum_{k\in\mathcal{K}_g} \rho_k^2\Big\{ \underbrace{ \|\bar{\boldsymbol g}_k - \tilde{\boldsymbol g}_k\|^2 + \frac{ K^\prime S_k  R_k}{N L \alpha_k^2} \sigma_{L,Q_k}^2
            }_{=\mathcal{E}_k(R_k)}
            \Big\}.
        \end{align}    
    \end{thm}
    \begin{IEEEproof}
        See Appendix~\ref{Apdx:MSE_upper}.
    \end{IEEEproof}  
    \vspace{1mm}
    {Theorem~\ref{Thm:MSE_upper}} shows that the upper bound of the global block reconstruction error at the PS can be minimized even if device $k\in\mathcal{K}_g$  locally minimizes $\mathcal{E}_k(R_k)$.
    Based on this result, instead of solving \eqref{eq:Prob_SRQ}, we formulate a {\em local} optimization problem for device $k\in\mathcal{K}_g$ as 
    \begin{align}
        R^{\star}_k &= \argmin_{R_k \in \mathcal{R}^{\rm c} }~ \|\bar{\boldsymbol g}_k - {\sf Sparse}(\bar{\boldsymbol g}_{k}) \|^2 + \frac{ K^\prime S_k R_k}{N L\alpha_k^2} \sigma_{L,Q_k}^2. \label{eq:Prob_SRQ2}
    \end{align}
    Because device $k$ knows the input local block update $\bar{\boldsymbol g}_k$ defined in \eqref{eq:true_local}, the objective function of \eqref{eq:Prob_SRQ2} can be computed for every $R_k \in \mathcal{R}^{\rm c}$ under the assumptions of $Q_k=Q_k^\star(R_k)$ and $S_k=S_k^\star (R_k)$.
    Then, by comparing the objective values, device $k$ can determine its best dimensionality reduction ratio $R^{\star}_k$.

    There exists a tradeoff between the first and second terms of the objective function in \eqref{eq:Prob_SRQ2} which can be interpreted as sparsification  and reconstruction errors, respectively. 
    The reconstruction error decreases with $R_k$ because the number of vector quantization bits $Q_k^\star (R_k)$ is an increasing function of $R_k$, as can be seen in \eqref{eq:Q}. 
    In contrast, the sparsification error increases with $R_k$ because the sparsity level $S_k^\star(R_k)$ is a decreasing function of $R_k$, as can be seen in \eqref{eq:S_2}. 
    Therefore, the optimization problem in \eqref{eq:Prob_SRQ2} allows each device to determine the best parameters for optimizing the tradeoff between the sparsification and reconstruction errors in FedVQCS.

    \subsection{Convergence Rate Analysis}\label{Sec:Convergence}
    In this analysis, we characterize the convergence rate of FedVQCS when the SGD algorithm with $E=1$ 
    is employed for updating the parameter vector at the PS.
    In our analysis, we make the following assumptions not only for mathematical tractability but also for connecting our analysis with the parameter optimization in Sec.~\ref{Sec:Param}.
    
    \vspace{1mm}
    {\bf Assumption 1:} 
    The loss function $F({\boldsymbol w})$ is $\beta$-smooth and is lower bounded by a constant $F({\boldsymbol w}^{\star})$, i.e.,  $F({\boldsymbol w})\geq F({\boldsymbol w}^{\star})$, $\forall {\boldsymbol w}\in \mathbb{R}^{\bar{N}}$. 
    
     {{\bf Assumption 2:}
    ~For every ${\boldsymbol w}$, the stochastic gradient is unbiased, i.e., $\mathbb{E}\big[\nabla F_k^{(t)}({\boldsymbol w})\big] = \nabla F_k({\boldsymbol w})$,  $\forall k,t$. Also, the variance of the stochastic gradient is bounded by a constant $\epsilon \geq 0$, i.e.,  $\mathbb{E}\big[\big\|\nabla F_k^{(t)}({\boldsymbol w}) - \nabla F_k({\boldsymbol w})\big\|^2\big] \leq \epsilon$,  $\forall k,t$. }

    {\bf Assumption 3:} 
    The squared sparsification error at the device $k$ is upper bounded by a constant $\frac{\zeta_b}{B} \geq 0$, i.e., $\mathbb{E}[\|\bar{\boldsymbol g}_k^{(t,b)} - \tilde{\boldsymbol g}_k^{(t,b)}\|^2] \leq 
    \frac{\zeta_b}{B}$, $\forall t,b$. 
    Also, the squared reconstruction error at the PS is upper bounded by the squared norm of the true block gradient with some constant $\delta_b < 1$, i.e., $\mathbb{E}[\|\tilde{\boldsymbol g}_\mathcal{K}^{(t,b)} - \hat{\boldsymbol g}_\mathcal{K}^{(t,b)}\|^2] \leq \delta_b\|\nabla F^{(b)}({\boldsymbol w}^{(t)})\|^2$, $\forall t,b$, where $\nabla F^{(b)}({\boldsymbol w}^{(t)}) \in \mathbb{R}^N$ is the $b$-th block of $\nabla F({\boldsymbol w}^{(t)})$.
    \vspace{1mm}
    
    {Assumption 3} is particularly relevant to FedVQCS because the parameter optimization of FedVQCS aims at reducing the MSE of the global block update by opportunistically determining the parameters $(Q,S,R)$ of the local block compression under the constraint of the wireless link capacity. 
    Note that Assumptions 1 and 2 are commonly adopted for analyzing the convergence rate of FL operating with a family of gradient descent algorithms, as in \cite{UVeqFed} and \cite{FedAvg_nonConvex}.

    Under {Assumptions 1--3}, we characterize the convergence rate of FedVQCS with the SGD algorithm.
    This result is given in the following theorem:
    \vspace{1mm}
     {\begin{thm}\label{Thm:Convergence}
        Under {Assumptions 1--3} with $E=1$, $\rho_k = \frac{1}{K}$, and $\gamma^{(t)} = \eta^{(t)} = \frac{1-\max_b\delta_b}{6\beta(1+\max_b\delta_b)\sqrt{T}}$, FedVQCS with the SGD algorithm satisfies the following bound:
        \begin{align}\label{eq:convergence}
            &\mathbb{E}\Bigg[\frac{1}{T}\sum_{t=1}^T\|\nabla F({\boldsymbol w}^{(t)})\|^2\Bigg] \nonumber \\ &\leq \frac{1}{\sqrt{T}}\Big[\tfrac{24\beta(1+\max_b\delta_b)}{(1-\max_b\delta_b)^2}\{F({\boldsymbol w}^{(1)}) - F({\boldsymbol w}^{\star})\} + \epsilon\Big] + \frac{\max_b\zeta_b}{18T}.
        \end{align}
    \end{thm}}
    \begin{IEEEproof}
        See Appendix~\ref{Apdx:Convergence}.
    \end{IEEEproof}
    \vspace{1mm}

    {Theorem~\ref{Thm:Convergence}} demonstrates that FedVQCS converges to a stationary point of the loss function if {Assumptions 1--3} hold. 
    This theorem also shows that the convergence rate of FedVQCS has the order of $\mathcal{O}\big(\frac{1}{\sqrt{T}}\big)$, which is the same as that of a classical FL framework for a non-convex loss function \cite{FedAvg_nonConvex}.
     {Furthermore, the convergence speed of FedVQCS is shown to improve as both $\delta_b$ and $\zeta_b$ decrease.}
    This result implies that our parameter optimization in Sec.~\ref{Sec:Param} is effective for improving the convergence speed. 

    \section{Simulation Results and Analysis}\label{Sec:Simul}
    In this section, we demonstrate the superiority of FedVQCS over existing FL frameworks, using simulations.
    We assume that the communication overhead allowed for transmitting the local model update at device $k$ is given by $C_k$ bits per local model entry.
	Under this assumption, we consider two communication scenarios: (i) {\em homogeneous} and (ii) {\em heterogeneous}.  In the homogeneous scenario, we set $C_k = \bar{C}$, $\forall k\in\mathcal{K}$, to model the wireless links with the same capacity; whereas, in the heterogeneous scenario, we uniformly draw $C_k$ from a pre-defined set $\mathcal{C}$ to model the wireless links with different capacities. 
    
     {In this simulation, we consider FL for an image classification task using the publicly accessible MNIST \cite{MNIST} and FEMNIST \cite{FEMNIST} datasets.} Details of the learning scenarios for each dataset are described below. 
    \begin{itemize}
        \item {\em MNIST:} 
        The global model is set to be a fully-connected neural network that consists of $784$ input nodes, a single hidden layer with $20$ hidden nodes, and $10$ output nodes. The activation functions of the hidden layer and the output layer are set to the rectified linear unit (ReLU) and the softmax function, respectively. 
        For the global model training at the PS, the ADAM optimizer in \cite{ADAM} with an initial learning rate $0.01$ is adopted. 
        Each local training dataset is determined by randomly selecting $500$ training data samples from two classes.
        For the local model training at each device, the mini-batch SGD algorithm with a learning rate $0.01$ is adopted with $|\mathcal{D}_k^{(t,e)}| = 10$ and $E = 3$.
	    For both the ADAM optimizer and the mini-batch SGD algorithm, the cross-entropy loss function is used.
	    For the MNIST dataset, we set the number of wireless devices to $K=75$ and the number of communication rounds to $T=50$. 
	     {\item {\em FEMNIST:}
	    The global model is set to be a variant ResNet-$18$, which has about $1.1\times10^7$ trainable parameters. For the original ResNet-$18$ in \cite{ResNet}, we modify the input, the first convolution, and output layers to match the dimensions of FEMNIST dataset.
	    We also replace the batch normalization layers to the layer normalization layers to improve the performance of the global model \cite{LN}.
	    For the global model training at the PS, the ADAM optimizer in \cite{ADAM} with an initial learning rate $5\times10^{-5}$ is adopted. For the local model training at each device, the mini-batch SGD algorithm with a learning rate $10^{-7}$ is adopted with $|\mathcal{D}_k^{(t,e)}| = 500$ and $E = 5$. 
	    Each local training dataset is determined by randomly grouping $20$ writers in FEMNIST dataset, and thus turn $200$ writers into $10$ devices \cite{FEMNIST1,FEMNIST2}.
	    For both the ADAM optimizer and the mini-batch SGD algorithm, the cross-entropy loss function is used. For the FEMNIST dataset, we set the number of wireless devices to $K=10$ and the number of communication rounds to $T=50$.}
	    
    \end{itemize}
	For performance comparisons, we consider the following FL baselines:
	\begin{itemize}
	    \item {\em Perfect reconstruction:} 
	    This framework assumes {\em lossless} transmission of the local model updates from the wireless devices to the PS without compression, as described in Sec.~\ref{Sec:FedAvg}. 

	    \item {\em FedVQCS:}  {FedVQCS is the proposed FL framework summarized in {\bf Procedure~\ref{alg:FedVQCS}}. In this framework, the vector quantizer in Sec.~\ref{Sec:VQDesign} is adopted. The parameters of FedVQCS are optimized as described in Sec.~\ref{Sec:Param} when the candidate set is $\mathcal{R}^{\rm c} = \big\{1.5 + 0.25r: r\in\{0,1,\ldots,6\}\big\}$. To be specific, the optimal dimensionality reduction ratio $R_k^\star$ is chosen among $\mathcal{R}^{\rm c}$ according to the criterion in \eqref{eq:Prob_SRQ2}, while the optimal quantization bit $Q_k^\star$ and the optimal sparsity level $S_k^\star$ are determined from \eqref{eq:Q} and \eqref{eq:S_2}, respectively.}
	    The EM-GAMP algorithm in \cite{FedQCS} is employed as a sparse signal recovery algorithm during the global block update reconstruction.
        The dimension of a subvector for vector quantization is set to be the largest integer, $L$, such that $L2^{Q_{\rm s}^\star} \leq 2^{15}$, where $Q_{\rm s}^\star$ is the optimal number of shape quantization bits.
	    
	    \item {\em ScalarQCS:} ScalarQCS is the communication-efficient FL framework developed in \cite{FedQCS} when $Q_k=2$, $R_k = 2/C_k$, and $S_k=\max\big(1,S_k^\star(R_k)\big)$.
	    The {\em aggregate-and-estimate} strategy in \cite{FedQCS} is adopted for the global block update reconstruction. 
	     
	    
	  


	    \item {\em HighVQ:} HighVQ is the communication-efficient FL framework developed in \cite{Du:20} when $Q_k=C_k$.
	    The dimension of each partition for vector quantization is set to be the largest integer, $L$, such that $L2^{C_k L} \leq 2^{15}$ for device $k$.
	    
	    \item {\em D-DSGD:} D-DSGD is the communication-efficient FL framework developed in \cite{Amiri:TSP}. 
	    The number of the nonzero entries for device $k$ is set to be $S_k$ such that $C_k\bar{N} = \log_2\big(\begin{smallmatrix} \bar{N} \\ S_k \end{smallmatrix}\big) + 33$.
        
	    
	\end{itemize}
	 {For FedVQCS and ScalarQCS, we set $(G,B)=(25,10)$ for the MNIST dataset and $(G,B)=(5,9152)$ for the FEMNIST dataset. Note that under this setting, the number of the devices in group is not larger than $3$, while $N$ is not larger than $2000$.}
	 {For  D-DSGD, we adjust the input data values of the MNIST training and test data samples so that these values are between $0$ and $1$. Except for this, we normalize the input data values of the training and test data samples using the mean and standard deviation of each dataset \cite{MNIST_normal_2}.}


	\begin{figure}
	    \vspace{-4mm}
		\centering 
		\subfigure[MNIST]
		{\epsfig{file=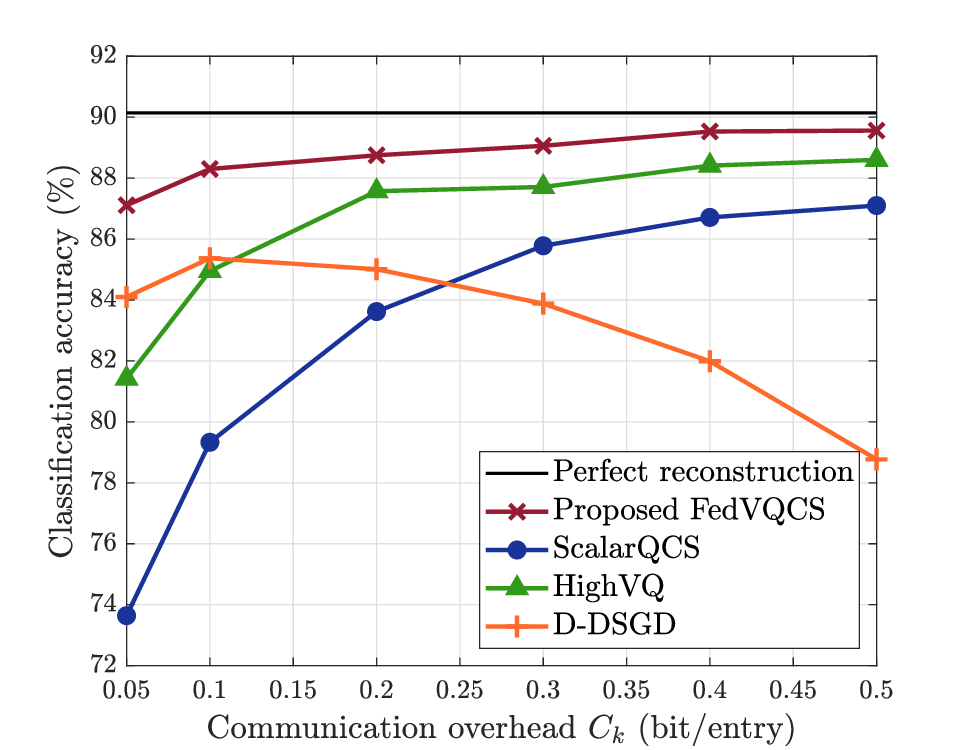, width=7.7cm}}
		\subfigure[FEMNIST]
		{\epsfig{file=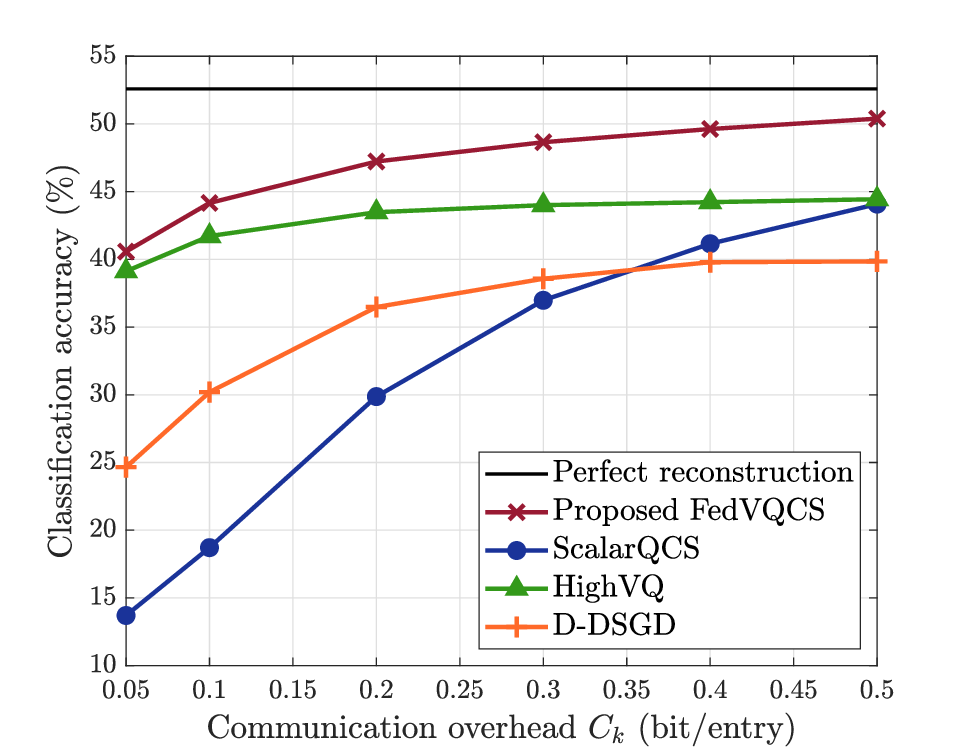, width=7.7cm}} \vspace{-2mm}
		\caption{Classification accuracy vs. communication overhead for different FL frameworks in the homogeneous scenario.} 
		\label{fig:Bitoverhead}
	\end{figure}
    

	In Fig.~\ref{fig:Bitoverhead}, we compare the classification accuracy of different FL frameworks in the homogeneous scenario.
	 {Fig.~\ref{fig:Bitoverhead} shows that for both the MNIST and FEMNIST datasets, FedVQCS provides more than a $2.4\%$ increase in classification accuracy compared to the existing communication-efficient FL frameworks  when the communication overhead of the local model update transmission is 0.1 bit per local model entry.} 
	Meanwhile, the results on the MNIST dataset demonstrate that FedVQCS with a 0.1-bit overhead per local model entry yields only a $2\%$ decrease in classification accuracy compared to the perfect reconstruction which requires a 32-bit overhead per local model entry for perfect transmission of the local model updates.
	This result demonstrates that FedVQCS enables an accurate reconstruction of the global model update at the PS, while significantly reducing the communication overhead of FL. 
	Fig.~\ref{fig:Bitoverhead} also shows that FedVQCS is more robust against the decrease in the communication overhead allowed for the devices, when compared with the existing FL frameworks.  

	\begin{figure}
		\centering 
		\subfigure[MNIST]
		{\hspace{-0.1cm}\epsfig{file=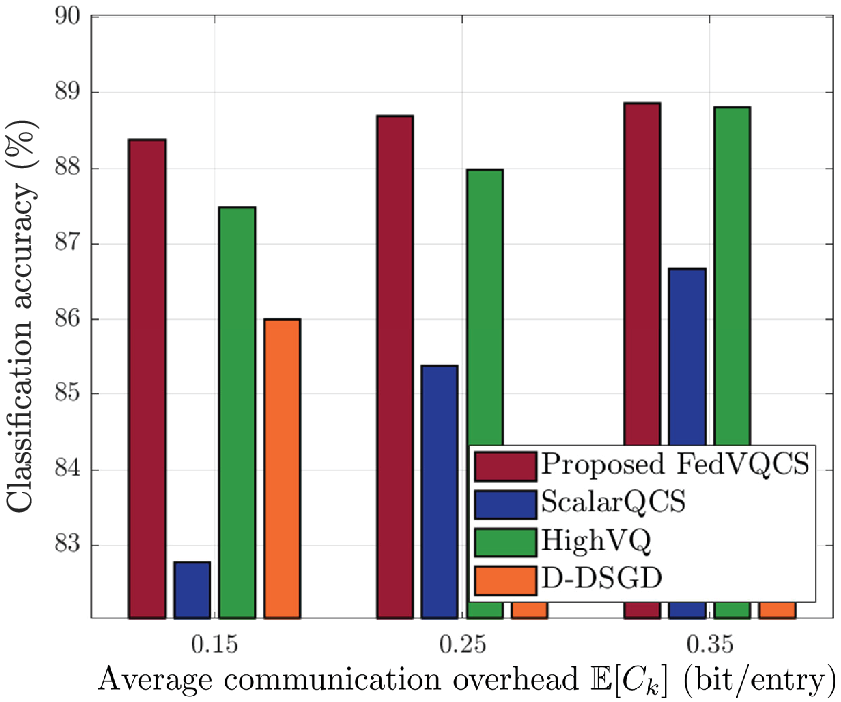, width=6.75cm}}
		\subfigure[FEMNIST]
		{\hspace{-0.1cm}\epsfig{file=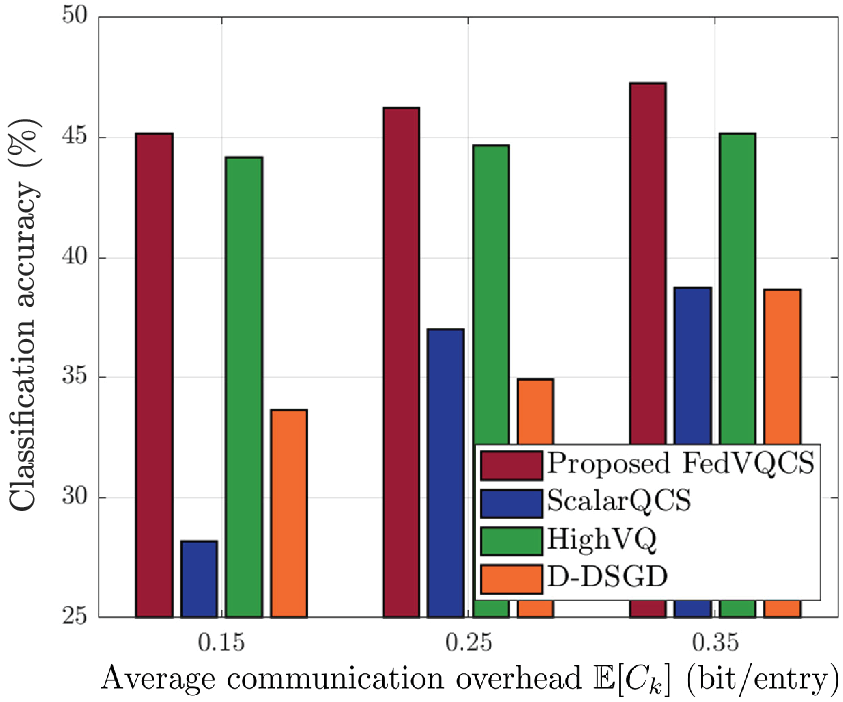, width=6.75cm}} \vspace{-2mm}
		\caption{Classification accuracy of different FL frameworks in the heterogeneous scenario with various average communication overheads.} 
		\label{fig:Hetero}
	\end{figure}

    In Fig.~\ref{fig:Hetero}, we compare the classification accuracy of different FL frameworks in the heterogeneous scenario with various average communication overheads. 
     {In this simulation, we set $\mathcal{C} = \{0.05,0.1,0.2,0.25\}$ for $\mathbb{E}[C_k] = 0.15$, $\mathcal{C} = \{0.05,0.15,0.35,0.45\}$ for $\mathbb{E}[C_k] = 0.25$,  and $\mathcal{C} = \{0.2,0.3,0.4,0.5\}$ for $\mathbb{E}[C_k] = 0.35$.}
    Fig.~\ref{fig:Hetero} shows that FedVQCS converges to the highest classification accuracy among the communication-efficient FL frameworks, irrespective of the average communication overhead allowed for the devices.
    These results imply that FedVQCS effectively adjusts the local model update compression at each device based on its link capacity, while reducing the  reconstruction error at the PS. 

\begin{figure}[t]
        \vspace{-3mm}
    	\centering
    	{\epsfig{file=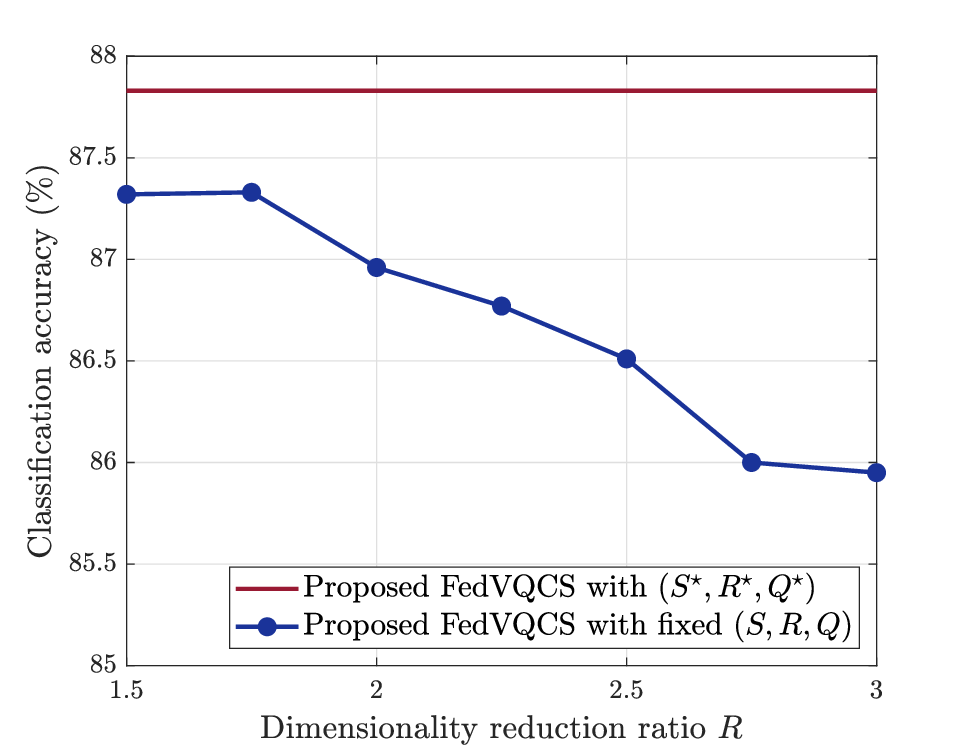, width=7.7cm}} 
    	\caption{Classification accuracy of FedVQCS with and without the parameter optimization for the MNIST dataset in the homogeneous scenario with $C_k = 0.07$.}  \vspace{-5mm}
    	\label{fig:Optimal_Fixed}
    \end{figure}

	In Fig.~\ref{fig:Optimal_Fixed}, we compare the classification accuracies of FedVQCS with and without the parameter optimization in Sec.~\ref{Sec:Param} for the MNIST dataset in the homogeneous scenario with $C_k = 0.07$.
	Without parameter optimization, we set $R_k=R$, $Q_k = Q_k^\star(R)$, and $S_k = S_k^\star(R)$, $\forall k \in\mathcal{K}$.
	Fig.~\ref{fig:Optimal_Fixed} shows that FedVQCS with the optimized parameters provides a $1.8\%$ increase in classification accuracy compared to FedVQCS with the {\em worst-case} parameters (i.e., $R=3$ case).
	This result demonstrates that our parameter optimization effectively improves the accuracy of global model reconstruction at the PS in FedVQCS. 
	This result is also consistent with the convergence rate analysis in Theorem \ref{Thm:Convergence} which demonstrated that the convergence rate improves as the reconstruction error of the global model update at the PS decreases.

    \begin{figure}[t]
    	\centering
    	{\epsfig{file=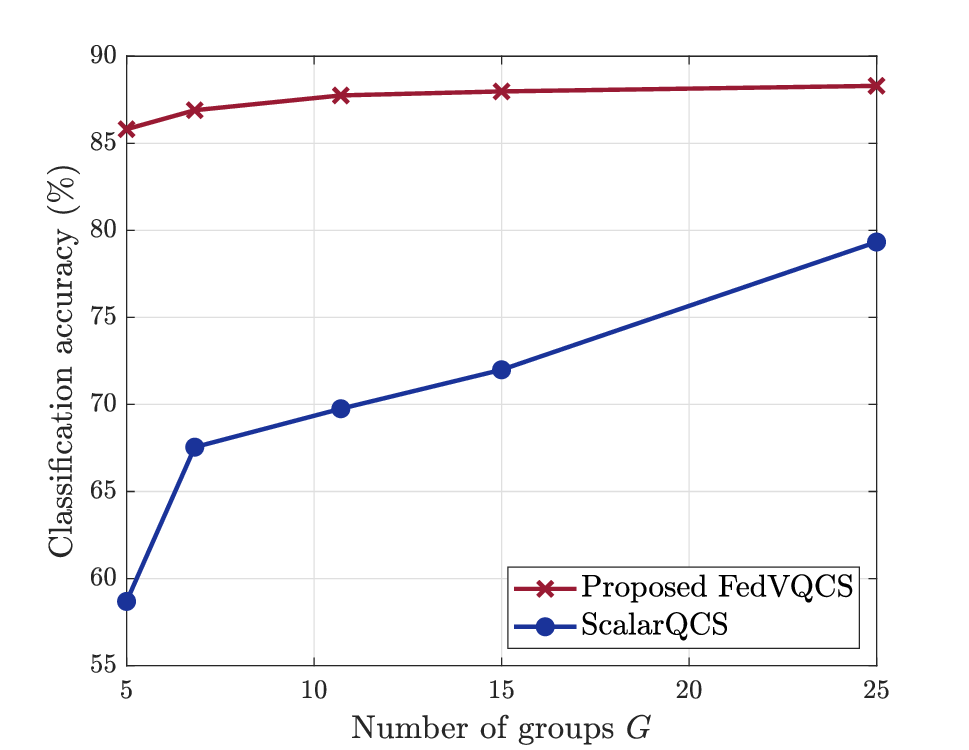, width=7.7cm}} 
    	\caption{Classification accuracy vs. number of groups for QCS-based FL frameworks for the MNIST dataset in the homogeneous scenario with $C_k = 0.1$.}  
    	\label{fig:Groups}
    \end{figure}

	In Fig.~\ref{fig:Groups}, we compare the classification accuracy of FedVQCS with that of ScalarQCS for different numbers of groups for the MNIST dataset in the homogeneous scenario with $C_k = 0.1$.
	Fig.~\ref{fig:Groups} shows that FedVQCS outperforms ScalarQCS regardless of the number of the groups.
    Because these two frameworks have a similar global model reconstruction process, the performance gain of FedVQCS over ScalarQCS is attributed to the use of vector quantization instead of scalar quantization.
    From Fig.~\ref{fig:Groups}, we can also observe that the classification accuracies of both frameworks decrease as $G$ decreases. 
    Because the computational complexity of the global model reconstruction process also decreases with $G$, the result in Fig.~\ref{fig:Groups} shows that the performance-complexity tradeoff associated with these frameworks can be controlled by adjusting the number of groups. 

	\section{Conclusion}
    In this paper, we have presented a novel framework for communication-efficient wireless FL. 
 	We have shown that our framework significantly reduces the communication overhead of FL by leveraging both vector quantization and CS-based dimensionality reduction for compressing the local model updates at the wireless devices.
 	Another key feature of our framework is that the devices can locally optimize the compression parameters in order to minimize the reconstruction error of a global model update at the PS under the constraint of the wireless link capacity. 
 	We have also demonstrated both analytically and numerically that our compression parameter optimization improves the convergence rate of FL.
 	An important direction of future research is to optimize the learning rate of a training algorithm employed at the PS by taking into account the reconstruction error at the PS. 
 	Another promising research direction is to develop dynamic device scheduling and  resource allocation strategies for the presented framework by considering the real-world radio resources available in a wireless network.

	\appendices
	\section{Proof of Lemma~\ref{lem:MSE_shape}}\label{Apdx:MSE_shape}
    For any even codebook, the following equality holds
    \begin{align}\label{eq:MSE_Cordal}
        \|{\boldsymbol s} - \hat{\boldsymbol s}\|^2 = 2-2\sqrt{1-d^2({\boldsymbol s}, \hat{\boldsymbol s})}.
    \end{align}
    In particular, for an {\em even} Grassmannian codebook $\mathcal{C}_{\rm s}^{+}$ of size $2^{Q_{\rm s}-1}$, the average chordal distance is bounded as 
    \begin{align}\label{eq:Cordal_shape}
        \frac{L-1}{L+1}2^{-\frac{2(Q_{\rm s}-1)}{L-1}} + o(1)
        \leq \mathbb{E}\left[ d^2( {\boldsymbol s}, \hat{\boldsymbol s}^{+}) \right] 
        \leq 2^{-\frac{2(Q_{\rm s}-1)}{L-1}} + o(1),
    \end{align}
    where $o(1)$ is a vanishing function of $L$ as $L\rightarrow\infty$ \cite{GrasslineAnalysis}.
    The bounds of the MSE in \eqref{eq:Cordal_shape} is also valid for $\mathbb{E}\big[ d^2( {\boldsymbol s}, \hat{\boldsymbol s}^{-}) \big]$, when $d^2({\boldsymbol s}, \hat{\boldsymbol s}^{+}) = d^2({\boldsymbol s}, \hat{\boldsymbol s}^{-})$.
    Because \eqref{eq:MSE_Cordal} is a convex function of $d^2({\boldsymbol s},\hat{\boldsymbol s})$, taking the expectation of \eqref{eq:MSE_Cordal} with respect to $d^2({\boldsymbol s},\hat{\boldsymbol s})$ and then applying Jensen's inequality with \eqref{eq:Cordal_shape} yields 
    \begin{align}\label{eq:MSE_low}
        2 - 2\sqrt{1-\frac{L-1}{L+1}2^{-\frac{2(Q_{\rm s}-1)}{L-1}} - o(1)}
        \leq
        \mathbb{E}\left[\| {\boldsymbol s} - \hat{\boldsymbol s}\|^2\right].
    \end{align}
    Based on the definition of the chordal distance, we have $\|{\boldsymbol s} - \hat{\boldsymbol s}\|^2 = 2-2\sqrt{1-d^2({\boldsymbol s}, \hat{\boldsymbol s})} = 2(1-|{\boldsymbol s}^{\sf T}\hat{\boldsymbol s}|)$ and
    $2d^2({\boldsymbol s}, \hat{\boldsymbol s}) = 2(1 + |{\boldsymbol s}^{\sf T}\hat{\boldsymbol s}|)(1 - |{\boldsymbol s}^{\sf T}\hat{\boldsymbol s}|)$.
    Because $1 + |{\boldsymbol s}^{\sf T}\hat{\boldsymbol s}| \geq 1$, we have $\|{\boldsymbol s} - \hat{\boldsymbol s}\|^2 \leq 2d^2({\boldsymbol s}, \hat{\boldsymbol s})$. Taking the expectation of this inequality and then applying the upper bound in \eqref{eq:Cordal_shape} yields 
    \begin{align}\label{eq:MSE_high}
        \mathbb{E}\left[\| {\boldsymbol s} - \hat{\boldsymbol s}\|^2\right]
        \leq        
        2^{-\frac{2(Q_{\rm s}-1)}{L-1} + 1} + 2o(1).
    \end{align}
    Combining the inequalities in \eqref{eq:MSE_low} and \eqref{eq:MSE_high} for a large $L$ yields the bounds in \eqref{eq:MSE_shape}.

    \section{Proof of Theorem~\ref{Thm:MSE_upper}}\label{Apdx:MSE_upper}
    In this proof, we assume that $\mathcal{K}_g = \{1,\ldots,K^\prime \}$ without loss of generality.
    Let $\mathcal{T} = \{i: (\tilde{\boldsymbol g}_{\mathcal{K}_g})_i\neq 0\}$ be the support set of $\tilde{\boldsymbol g}_{\mathcal{K}_g}$.
    Also, let $R_g$ be the dimensionality reduction ratio of the devices in group $g$.
    With the knowledge of $\mathcal{T}$, the objective function in \eqref{eq:Prob_SRQ} can be rewritten as 
    \begin{align}\label{eq:MSE_prime}
        \mathbb{E}[\|\bar{\boldsymbol g}_{\mathcal{K}_g} - \hat{\boldsymbol g}_{\mathcal{K}_g}  \|^2 ]  
        &=\|\bar{\boldsymbol g}_{\mathcal{K}_g} - \tilde{\boldsymbol g}_{\mathcal{K}_g}\|^2 + \mathbb{E}[\|\tilde{\boldsymbol g}_{\mathcal{K}_g} - \hat{\boldsymbol g}_{\mathcal{K}_g}\|^2].
    \end{align}
    The first term of the right-hand-side (RHS) of \eqref{eq:MSE_prime} is rewritten as
    \begin{align}\label{eq:MSE_first}
        \|\bar{\boldsymbol g}_{\mathcal{K}_g} - \tilde{\boldsymbol g}_{\mathcal{K}_g}\|^2
        &= \bigg\| \sum_{k\in\mathcal{K}_g}\rho_k^2\{ \bar{\boldsymbol g}_k - \tilde{\boldsymbol g}_k\} \bigg\|^2 
        \nonumber \\&\overset{(a)}{\leq} 
        K^\prime \sum_{k\in\mathcal{K}_g}\rho_k^2\|\bar{\boldsymbol g}_k - \tilde{\boldsymbol g}_k\|^2,
    \end{align}
    where $(a)$ follows from the Cauchy–Schwarz inequality.
    Let $\tilde{\boldsymbol g}_{\mathcal{T}}$ be a subvector of $\tilde{\boldsymbol g}_{\mathcal{K}_g}$, which consists of the entries of  $\tilde{\boldsymbol g}_{\mathcal{K}_g}$ whose indices belong to $\mathcal{T}$.
    Similarly, let ${\boldsymbol A}_{\mathcal{T}}$ be a submatrix of ${\boldsymbol A}_{R_g}$, which consists of the columns of ${\boldsymbol A}_{R_g}$ whose column indices belong to $\mathcal{T}$.
    Then, with the knowledge of $\mathcal{T}$, the sparse signal recovery problem in \eqref{eq:aggregate_FedVQCS} reduces to the estimation of $\tilde{\boldsymbol g}_{\mathcal{T}}$ from its linear observation ${\boldsymbol y}_{\mathcal{K}_g} = {\boldsymbol A}_{R_g}\tilde{\boldsymbol g}_{\mathcal{K}_g} + {\boldsymbol d}_{\mathcal{K}_g} = {\boldsymbol A}_{\mathcal{T}}\tilde{\boldsymbol g}_{\mathcal{T}} + {\boldsymbol d}_{\mathcal{K}_g}$. The least squares solution of this problem is given by 
        \begin{align}
            \hat{\boldsymbol g}_{\mathcal{T}} = \tilde{\boldsymbol g}_{\mathcal{T}} + \big({\boldsymbol A}_{\mathcal{T}}^{\sf T}{\boldsymbol A}_{\mathcal{T}}\big)^{-1}{\boldsymbol A}_{\mathcal{T}}^{\sf T}{\boldsymbol d}_{\mathcal{K}_g}.
        \end{align} 
    Therefore, the second term of the RHS of \eqref{eq:MSE_prime} is computed as 
    \begin{align}
        &\mathbb{E}[\|\tilde{\boldsymbol g}_{\mathcal{K}_g} - \hat{\boldsymbol g}_{\mathcal{K}_g}\|^2] 
        = \mathbb{E}[\|\tilde{\boldsymbol g}_{\mathcal{T}} - \hat{\boldsymbol g}_{\mathcal{T}}\|^2] \nonumber \\ &= \mathbb{E}\big[\big\|\big({\boldsymbol A}_{\mathcal{T}}^{\sf T}{\boldsymbol A}_{\mathcal{T}}\big)^{-1}{\boldsymbol A}_{\mathcal{T}}^{\sf T}{\boldsymbol d}_{\mathcal{K}_g}\big\|^2\big] \nonumber \\
        &= {\sf Tr}\big(\big({\boldsymbol A}_{\mathcal{T}}^{\sf T}{\boldsymbol A}_{\mathcal{T}}\big)^{-1} {\boldsymbol A}_{\mathcal{T}}^{\sf T} \mathbb{E}[ {\boldsymbol d}_{\mathcal{K}_g}{\boldsymbol d}_{\mathcal{K}_g}^{\sf T}] {\boldsymbol A}_{\mathcal{T}}\big({\boldsymbol A}_{\mathcal{T}}^{\sf T}{\boldsymbol A}_{\mathcal{T}}\big)^{-1}  \big)  \nonumber \\
        &\overset{(a)}{=} \frac{\tilde{S}_g }{M_k^2}\mathbb{E}[ \|{\boldsymbol d}_{\mathcal{K}_g}\|^2] \overset{(b)}{\leq} \frac{K^\prime S_k }{M_k^2}\mathbb{E}[ \|{\boldsymbol d}_{\mathcal{K}_g}\|^2],
        \label{eq:MSE_second}
    \end{align}    
    where $(a)$ holds when $\frac{N}{R_k} = M_k \gg 1$ and $\tilde{S}_g \gg 1$ because ${\boldsymbol A}_{\mathcal{T}}^{\sf T}{\boldsymbol A}_{\mathcal{T}} \rightarrow M_k{\boldsymbol I}_{\tilde{S}_g}$ as $M_k \rightarrow \infty$ and ${\boldsymbol A}_{\mathcal{T}}{\boldsymbol A}_{\mathcal{T}}^{\sf T} \rightarrow \tilde{S}_g {\boldsymbol I}_{M_k}$ as $\tilde{S}_g \rightarrow \infty$ by the law of large numbers, and $(b)$ holds because $\tilde{S}_g \leq K^\prime S_k$ from our discussion on \eqref{eq:S_2}. 
    The quantization error $\mathbb{E}[ \|{\boldsymbol d}_{\mathcal{K}_g}\|^2]$ in \eqref{eq:MSE_second} is upper bounded as
    \begin{align}\label{eq:MSE_d_Kg}
        \mathbb{E}[ \|{\boldsymbol d}_{\mathcal{K}_g}\|^2]
        &\overset{(a)}{\leq} 
        K^\prime \sum_{k\in\mathcal{K}_g}\frac{\rho_k^2}{\alpha_k^2} \mathbb{E}[ \left\|{\boldsymbol d}_k\right\|^2] \nonumber \\ &\overset{(b)}{=}   K^\prime \frac{M_k}{L} \sum_{k\in\mathcal{K}_g}\frac{\rho_k^2}{\alpha_k^2}  \sigma_{L,Q_k}^2, 
    \end{align}    
    where $(a)$ follows from the Cauchy–Schwarz inequality, and $(b)$ follows from \eqref{eq:MSE_VQ}.
    Applying \eqref{eq:MSE_first}--\eqref{eq:MSE_d_Kg} into \eqref{eq:MSE_prime} yields the result in \eqref{eq:MSE_upper}.

    \section{Proof of Theorem~\ref{Thm:Convergence}}\label{Apdx:Convergence}
	
	In this proof, we consider a virtual sequence of parameter vectors, defined as $\tilde{\boldsymbol w}^{(t+1)} = \tilde{\boldsymbol w}^{(t)} - \eta \{\nabla F^{(t)}({\boldsymbol w}^{(t)}) - {\boldsymbol e}^{(t)}\}$, where $\nabla F^{(t)}({\boldsymbol w}^{(t)}) = \frac{1}{K}\sum_{k}\nabla F_k^{(t)}\big({\boldsymbol w}^{(t)}\big)$, ${\boldsymbol e}^{(t)} = \tilde{\boldsymbol g}_{\mathcal{K}}^{(t)} - \hat{\boldsymbol g}_{\mathcal{K}}^{(t)}$ and $\tilde{\boldsymbol w}^{(1)} = {\boldsymbol w}^{(1)}$. 
	Then, by the definition in \eqref{eq:true_local}, we have $\tilde{\boldsymbol w}^{(t)} = {\boldsymbol w}^{(t)} - \frac{\eta }{K}\sum_k{\boldsymbol \Delta}_k^{(t-1)}$, $\forall t>1$.
    Under Assumption~1, the improvement of the loss function in round $t$ satisfies
    \begin{align}
        &\mathbb{E}[F(\tilde{\boldsymbol w}^{(t+1)}) - F(\tilde{\boldsymbol w}^{(t)})]
        \nonumber \\& \leq
        \mathbb{E}[\nabla F(\tilde{\boldsymbol w}^{(t)})^{\sf T}(\tilde{\boldsymbol w}^{(t+1)}-\tilde{\boldsymbol w}^{(t)})] + \frac{\beta}{2}\mathbb{E}[\|\tilde{\boldsymbol w}^{(t+1)}-\tilde{\boldsymbol w}^{(t)}\|^2]
        \nonumber \\ &\overset{(a)}{=} 
        -{\eta} \mathbb{E}[\nabla F(\tilde{\boldsymbol w}^{(t)})^{\sf T}\nabla F({\boldsymbol w}^{(t)})] + {\eta} \mathbb{E}[\nabla F(\tilde{\boldsymbol w}^{(t)})^{\sf T}{\boldsymbol e}^{(t)}] \nonumber \\ &~~~~+ \frac{{\eta}^2 \beta}{2}\mathbb{E}[\|\nabla F^{(t)}({\boldsymbol w}^{(t)}) - {\boldsymbol e}^{(t)}\|^2],
        \label{eq:loss_improve}
    \end{align}
    where the expectation is taken over the randomness in the trajectory, and $(a)$ follows from the definition of $\tilde{\boldsymbol w}^{(t)}$ with Assumption 2. 
    Note that $2\nabla F(\tilde{\boldsymbol w}^{(t)})^{\sf T}\nabla F({\boldsymbol w}^{(t)}) = \|\nabla F(\tilde{\boldsymbol w}^{(t)})\|^2 + \|\nabla F({\boldsymbol w}^{(t)})\|^2 - \|\nabla F(\tilde{\boldsymbol w}^{(t)}) - \nabla F({\boldsymbol w}^{(t)})\|^2$.
     {In addition, under Assumptions 2 and 3, we have 
    \begin{align}
        &\mathbb{E}[ \nabla F(\tilde{\boldsymbol w}^{(t)})^{\sf T}{\boldsymbol e}^{(t)} ]
        \leq \frac{\mathbb{E}[\|\nabla F(\tilde{\boldsymbol w}^{(t)})\|^2 + \|{\boldsymbol e}^{(t)}\|^2]}{2} 
        \nonumber \\&\qquad\qquad~~\leq \frac{\mathbb{E}[\|\nabla F(\tilde{\boldsymbol w}^{(t)})\|^2]}{2} + \frac{\max_b\delta_b }{2} \mathbb{E}[\|\nabla F({\boldsymbol w}^{(t)})\|^2], \nonumber \\
        &\mathbb{E}[\|\nabla F^{(t)}({\boldsymbol w}^{(t)}) - {\boldsymbol e}^{(t)}\|^2] 
        \nonumber \\&=  \mathbb{E}[\|\nabla F^{(t)}({\boldsymbol w}^{(t)}) - \nabla F({\boldsymbol w}^{(t)}) + \nabla F({\boldsymbol w}^{(t)}) - {\boldsymbol e}^{(t)}\|^2] \nonumber \\
        &\leq 3\mathbb{E}\bigg[\frac{1}{K}\sum_{k=1}^K\|\nabla F_k^{(t)}({\boldsymbol w}^{(t)}) - \nabla F_k({\boldsymbol w}^{(t)})\|^2 
        \nonumber \\ &\qquad~~~ + \|\nabla F({\boldsymbol w}^{(t)})\|^2 + \|{\boldsymbol e}^{(t)}\|^2\bigg]
        \nonumber \\
        &\leq 3[\epsilon + (1 + \max_b\delta_b) \mathbb{E}[\|\nabla F({\boldsymbol w}^{(t)})\|^2]].
    \end{align}
    Applying the above inequalities into \eqref{eq:loss_improve} yields
    \begin{align} \label{eq:loss_improve2}
        &\mathbb{E}[F(\tilde{\boldsymbol w}^{(t+1)}) - F(\tilde{\boldsymbol w}^{(t)})]
        \nonumber \\ &\!\!\leq 
        - \frac{{\eta} \bar{\delta}}{2}\mathbb{E}[\|\nabla F({\boldsymbol w}^{(t)})\|^2] + \frac{{\eta}}{2}\mathbb{E}[\|\nabla F(\tilde{\boldsymbol w}^{(t)}) - \nabla F({\boldsymbol w}^{(t)})\|^2] \nonumber \\ &~~~~ + \frac{3{\eta}^2\beta\epsilon}{2}
        \nonumber \\&\!\!\overset{(a)}{\leq} 
        - \frac{{\eta}\bar{\delta}}{2}\mathbb{E}[\|\nabla F({\boldsymbol w}^{(t)})\|^2] + \frac{{\eta}\beta^2}{2}\mathbb{E}[\|\tilde{\boldsymbol w}^{(t)} - {\boldsymbol w}^{(t)}\|^2]+ \frac{3{\eta}^2\beta\epsilon}{2},
    \end{align}
    where $\bar{\delta} = 1-\max_b\delta_b-3{\eta}\beta(1+\max_b\delta_b)$, and $(a)$ follows from Assumption 1 implying that $\|\nabla F(\tilde{\boldsymbol w}^{(t)}) - \nabla F({\boldsymbol w}^{(t)})\|^2 \leq \beta^2\|\tilde{\boldsymbol w}^{(t)} - {\boldsymbol w}^{(t)}\|^2$.
    Under Assumption 2, we also have $\mathbb{E}[\|\tilde{\boldsymbol w}^{(t)} - {\boldsymbol w}^{(t)}\|^2] \leq \frac{\eta^2}{K}\sum_k\sum_b\mathbb{E}[\|{\boldsymbol \Delta}_k^{(t-1, b)}\|^2] \leq        
        \eta^2\max_b\zeta_b$.
    Applying the above inequality into \eqref{eq:loss_improve2} with ${\eta} = \frac{1-\max_b\delta_b}{6\beta(1+\max_b\delta_b)\sqrt{T}}$ yields
    \begin{align}
        &\mathbb{E}[F(\tilde{\boldsymbol w}^{(t+1)}) - F(\tilde{\boldsymbol w}^{(t)})]
        \nonumber \\
        &\leq
        - \frac{(1-\max_b\delta_b)^2}{12\beta(1+\max_b\delta_b)\sqrt{T}}\bigg(1 - \frac{1}{2\sqrt{T}}\bigg)\mathbb{E}[\|\nabla F({\boldsymbol w}^{(t)})\|^2] \nonumber \\
        &~~~ +\frac{(1-\max_b\delta_b)^3\max_b\zeta_b}{432\beta(1+\max_b\delta_b)^3 T\sqrt{T}}  + \frac{(1-\max_b\delta_b)^2\epsilon}{24\beta(1+\max_b\delta_b)^2 T}
        \nonumber \\ 
        &\overset{(a)}{\leq}
        - \frac{(1-\max_b\delta_b)^2}{24\beta(1+\max_b\delta_b)}\bigg[\frac{1}{\sqrt{T}}\mathbb{E}[\|\nabla F({\boldsymbol w}^{(t)})\|^2] \nonumber \\ &\qquad\qquad\qquad\qquad\qquad~~~ - \frac{\max_b\zeta_b}{18T\sqrt{T}} - \frac{\epsilon}{T}\bigg],
        \label{eq:loss_improve4}
    \end{align}
    where $(a)$ follows from $\frac{1}{2\sqrt{T}}\leq\frac{1}{2}$, $\frac{(1-\max_b\delta_b)^3}{(1+\max_b\delta_b)^3} \leq \frac{(1-\max_b\delta_b)^2}{1+\max_b\delta_b}$, and $\frac{(1-\max_b\delta_b)^2}{(1+\max_b\delta_b)^2} \leq \frac{(1-\max_b\delta_b)^2}{1+\max_b\delta_b}$ for $0\leq \delta_b < 1$, $\forall b$.
    By considering the telescoping sum over the iterations, a lower bound of the initial loss with ${\boldsymbol w}^{(1)}$ is expressed as
    \begin{align}
        &F({\boldsymbol w}^{(1)}) - F({\boldsymbol w}^{\star}) \geq F({\boldsymbol w}^{(1)}) - \mathbb{E}[F(\tilde{\boldsymbol w}^{T+1})] \nonumber \\ &= \sum_{t=1}^T \mathbb{E}[F(\tilde{\boldsymbol w}^{(t)}) - F(\tilde{\boldsymbol w}^{(t+1)})]
        \nonumber \\ 
        &\geq
        \frac{(1-\max_b\delta_b)^2}{24\beta(1+\max_b\delta_b)}\bigg[\frac{1}{\sqrt{T}}\sum_{t=1}^T\mathbb{E}[\|\nabla F({\boldsymbol w}^{(t)})\|^2] \nonumber \\ &\qquad\qquad\qquad~~~~~~~~~~ - \frac{\max_b\zeta_b}{18\sqrt{T}} - \epsilon\bigg].
        \label{eq:loss_improve5}
    \end{align}
    The inequality in \eqref{eq:loss_improve5} can be rewritten as in \eqref{eq:convergence}, which completes the proof. }

\end{document}